\documentclass[%
 reprint,
 superscriptaddress,
 preprintnumbers,
 nofootinbib,
 amsmath,amssymb,
 aps,
 prl,
]{revtex4-2}

\usepackage{algorithm}
\usepackage{algpseudocode}
\usepackage{psfrag,slashed,cancel,array,graphicx,todonotes}
\usepackage{subfigure}
\usepackage[labelfont=bf,labelsep=quad,justification=justified]{caption}
\usepackage[utf8]{inputenc}
\usepackage{mathtools}
\usepackage{mathrsfs}
\usepackage[mathscr]{euscript}
\usepackage{multirow}
\usepackage{braket}
\usepackage{slashed}
\usepackage{booktabs} 
\usepackage[normalem]{ulem}
\usepackage{slashed}
\usepackage{hyperref}

\hypersetup{pdftitle={RG evolution of the B-meson LCDA and radiative jet function at three loops},pdfcreator={},linkcolor=[rgb]{0.15,0.35,0.75},colorlinks=true,citecolor=[rgb]{0.675,0,0.2},urlcolor=[rgb]{0.15,0.35,0.65}}

\newcommand{\bn}{\bar{n}}
\newcommand{\nn}{\nonumber}

\newcommand{\cO}{{\mathcal O}}

\newcommand{\cM}{{\mathcal M}}

\newcommand{\cS}{{\mathcal S}}
\newcommand{\cJ}{{\mathcal J}}

\begin{document}

\title{RG evolution of the $B$-meson LCDA and radiative jet function at three loops}


\author{Ze Long Liu}\email{liuzelong@ihep.ac.cn}
\author{Guang Tang}\email{tangg@ihep.ac.cn}
\author{Zhe-Yu Wang}\email{wangzheyu@ihep.ac.cn}

\affiliation{ Theoretical Physics Division, Institute of High Energy Physics, Chinese Academy of Sciences, Beijing 100049, China}

\affiliation{School of Physical Sciences, University of Chinese Academy of Sciences, Beijing 100049, China}

\begin{abstract}
We calculate the three-loop evolution kernel of the light-cone distribution amplitude of the $B$ meson, together with the radiative jet function in the leptonic radiative decay $B^-\to\gamma\ell^-\bar\nu$ for the first time. Representing the light-cone distribution amplitude as the discontinuity of a matrix element containing an inverse differential operator, we compute the partonic matrix element in momentum space with a novel method of infrared subtraction. We show that its infrared singularities are governed by the same anomalous dimension as that of the heavy-light quark form factor, so that they can be subtracted without introducing any infrared regulator, thereby preserving gauge invariance and avoiding additional complexity in the Feynman integrals. Based on the renormalization-group invariance of the factorization theorem, we derive the renormalization factor of the radiative jet function, and confirm that it removes all divergences of the bare three-loop radiative jet function. Our results enable renormalization-group improved predictions for exclusive $B$ decays at next-to-next-to-next-to-leading order.
\end{abstract}

\maketitle

\paragraph*{Introduction.---}
The light-cone distribution amplitude (LCDA) $\phi_+^B(\omega,\mu)$ of the $B$ meson, defined by a bilocal heavy-light operator in heavy-quark effective theory (HQET)~\cite{Grozin:1996pq}, is an essential ingredient in theoretical predictions of exclusive $B$ decays.
Based on QCD factorization~\cite{Beneke:1999br,Beneke:2000ry} and soft-collinear effective theory (SCET)~\cite{Bauer:2000ew,Bauer:2000yr,Bauer:2001ct,Bauer:2001yt,Beneke:2002ph},
the amplitudes of many exclusive $B$ decays, such as $B\to\pi\pi$, $B\to K^*\gamma$ and $B\to\gamma\ell\nu$, can be expressed at leading power in $\Lambda_{\rm QCD}/m_b$ as convolutions of the form~\cite{Lange:2003ff} 
\begin{equation}\label{eq:conv}
\int_0^\infty\frac{d\omega}{\omega}\,T(\omega,\mu)\,\phi_+^B(\omega,\mu)\,,
\end{equation}
where $T$ is a perturbative hard-scattering kernel. For the leptonic radiative decay $B^-\to\gamma\ell^-\bar\nu$, the amplitude is described by the factorization theorem~\cite{Lunghi:2002ju,Descotes-Genon:2002crx,Bosch:2003fc}
\begin{align}\label{eq:Bgamma}
\cM(B^-\!\to\gamma\ell^-\bar\nu)&\propto F(\mu)\,H(\mu) \nn\\
&\times\int_0^\infty\frac{d\omega}{\omega}\,J(-2E_\gamma\omega,\mu)\,\phi_+^B(\omega,\mu)\,,
\end{align}
where $F(\mu)$ is the HQET decay constant, $H$ is a hard function depending on the $b$-quark mass $m_b$ and the photon energy $E_\gamma$, and $J$ is the radiative jet function at the hard-collinear scale $2E_\gamma\omega\sim m_b\Lambda_{\rm QCD}$. This decay can be used to probe the inverse moment $\lambda_B$ of the LCDA directly~\cite{Beneke:2011nf,Belle:2018jqd}, and its power corrections have been studied extensively~\cite{Braun:2012kp,Wang:2016qii,Beneke:2018wjp,Wang:2018wfj}. Large logarithms of the ratios of the hard, hard-collinear and soft scales are resummed by solving the renormalization-group (RG) equations of the functions in~\eqref{eq:Bgamma}. With the hard and jet functions and the anomalous dimensions at two loops~\cite{Asatrian:2008uk,Bonciani:2008wf,Bell:2008ws,Beneke:2008ei,Bell:2010mg,Braun:2019wyx,Liu:2020ydl}, the amplitude is known up to next-to-next-to-leading order in RG-improved perturbation theory~\cite{Korchemsky:1999qb,Descotes-Genon:2002crx,Lunghi:2002ju,Bosch:2003fc,Beneke:2011nf,Galda:2020epp,Galda:2022dhp}. To explore exclusive $B$ decays at higher precision, the main challenge comes from the three-loop evolution kernel of $\phi_+^B(\omega,\mu)$. The kernel is also required to evolve the LCDA from a hadronic scale, where it is constrained by QCD sum rules~\cite{Braun:2003wx}, Euclidean lattice quantities~\cite{Wang:2019msf} or the matching onto boosted heavy mesons~\cite{Beneke:2023nmj}, to the hard-collinear scale.

The one-loop evolution kernel, known as the Lange-Neubert kernel, was derived in~\cite{Lange:2003ff}, where the coefficient of the logarithm of the renormalization scale was identified with the cusp anomalous dimension~\cite{Korchemsky:1987wg,Braun:2003wx}. The one-loop evolution is local in a dual space~\cite{Bell:2013tfa}, and the kernel is fixed by conformal symmetry along the light cone~\cite{Braun:2014owa}, which was used to derive the two-loop kernel~\cite{Braun:2019wyx}. To compute the three-loop kernel directly from the partonic matrix element, one first needs to isolate the ultraviolet (UV) divergences from the infrared (IR) singularities. This is usually achieved by introducing an internal mass or off-shellness as an additional IR regulator~\cite{Descotes-Genon:2002crx,Grozin:2005iz}. However, this would extremely complicate the Feynman integrals at three loops. 

In this {\it Letter}, we present the first calculation of the evolution kernel of the $B$-meson LCDA and the radiative jet function at three loops. We represent the LCDA as the discontinuity of a matrix element containing an inverse differential operator in momentum space, which makes integration-by-parts (IBP) reduction and differential equations (DEs) applicable. We demonstrate that the IR singularities of the on-shell partonic matrix elements are governed by the same anomalous dimension as that of the heavy-light quark form factor, so that they can be subtracted without introducing any additional IR regulator. We also calculate the three-loop radiative jet function independently following the method in~\cite{Liu:2020ydl}. Deriving its renormalization factor from the RG invariance of~\eqref{eq:Bgamma}, we confirm that it cancels all poles of the bare jet function at three loops. 
Finally, we present the RG-improved predictions for the partial branching fraction of $B^-\to\gamma\ell^-\bar\nu$ up to next-to-next-to-next-to-leading order (N$^3$LO) in QCD.

\paragraph*{$B$-meson LCDA and radiative jet function in momentum space.---}
The $B$-meson LCDA is defined by the Fourier transform of a bilocal HQET matrix element~\cite{Grozin:1996pq,Lange:2003ff},
\begin{align}\label{eq:LCDAdef}
\phi_+^B(\omega,\mu)&=\frac{1}{2\pi i\, F(\mu)}\int d\tau\,e^{i\omega\tau} \nn\\
&\quad\times\langle 0|\,\bar q_s(\tau n)\,S_n(\tau n,0)\,\slashed{n}\gamma_5\,h_v(0)\,|\bar B(v)\rangle\,,
\end{align}
where $h_v$ is the heavy-quark field in HQET, $q_s$ is the soft spectator quark, $v$ is the $B$-meson velocity, $n$ is a light-like vector satisfying $n\cdot v=1$, and $S_n(\tau n,0)$ is a finite-length soft Wilson line ensuring the gauge invariance of the matrix element. The matrix element is understood as a function of $\tau-i0$, such that $\phi_+^B(\omega,\mu)$ vanishes for $\omega<0$~\cite{Lange:2003ff}. The antiquark field together with the Wilson line can be translated to the origin by $\bar q_s(\tau n)\,S_n(\tau n,0)=\bar q_s(0)\,e^{\tau\,n\cdot\overleftarrow{D}}$, where $\overleftarrow{D}$ is the covariant derivative acting on $\bar q_s$. After integrating over $\tau$ in~\eqref{eq:LCDAdef}, the LCDA can be rewritten as
\begin{equation}\label{eq:local}
\phi_+^B(\omega,\mu)=\frac{1}{i F(\mu)}\,
\langle 0|\,O_+(\omega)\,|\bar B(v)\rangle\,,
\end{equation}
with
\begin{equation}\label{eq:Oplus}
O_+(\omega) = \bar q_s(0)\,\delta\big(\omega-in\cdot\overleftarrow{D}\big)\,\slashed{n}\gamma_5\,h_v(0)\,.
\end{equation}
The non-local operator in position space has thus been converted into a product of two fields at the origin, separated by a differential operator that measures the light-cone momentum of the spectator antiquark. The LCDA obeys the RG equation~\cite{Lange:2003ff}
\begin{equation}\label{eq:RGE}
\frac{d}{d\ln\mu}\,\phi_+^B(\omega,\mu)=-\int_0^\infty d\omega'\,\gamma_+(\omega,\omega',\mu)\,\phi_+^B(\omega',\mu)\,,
\end{equation}
where the evolution kernel $\gamma_+(\omega,\omega',\mu)$ is determined by the renormalization of $O_+(\omega)$~\cite{Lange:2003ff}
\begin{equation}\label{eq:Zplus}
O_+^{\rm ren}(\omega,\mu)=\int d\omega'\,Z_+(\omega,\omega',\mu)\,O_+^{\rm bare}(\omega')\,.
\end{equation}
The renormalization factor $Z_+$ absorbs the UV poles of the matrix elements of $O_+^{\rm bare}$ in the $\overline{\rm MS}$ scheme, and the evolution kernel is given by~\cite{Lange:2003ff}
\begin{align}\label{eq:gamma}
\gamma_+(\omega,\omega',\mu)&=-\int d\tilde\omega\,\frac{dZ_+(\omega,\tilde\omega,\mu)}{d\ln\mu}\,Z_+^{-1}(\tilde\omega,\omega',\mu) \nn\\
&\quad-\gamma_F(\alpha_s)\,\delta(\omega-\omega')\,,
\end{align}
where $\gamma_F$ is the anomalous dimension of $F(\mu)$, known to three-loop order~\cite{Broadhurst:1991fz,Chetyrkin:2003vi}.

To perform the calculation conveniently in momentum space, we convert the $\delta$-function into a ``propagator'' by using reverse unitarity~\cite{Anastasiou:2002yz},
\begin{equation}\label{eq:discO}
\begin{aligned}
O_+(\omega)=&-\frac{1}{2\pi}\,{\rm Disc}_\omega\,\cO_+(\omega)\\
=&-\frac{1}{2\pi}\,{\rm Disc}_\omega\left[
\bar q_s(0)\,\frac{i}{in\cdot\overleftarrow{D}-\omega}\,\slashed{n}\gamma_5\,h_v(0)\right]\,,
\end{aligned}
\end{equation}
with ${\rm Disc}_\omega f(\omega)\equiv f(\omega+i0)-f(\omega-i0)$. The $\delta$-function and the plus distributions in $Z_+(\omega,\omega',\mu)$ are recovered by taking the discontinuity in $\omega$ at the end.

To obtain $Z_+(\omega,\omega',\mu)$, we evaluate the matrix element of $\cO_+(\omega)$ between the vacuum and an on-shell partonic state consisting of a heavy quark with velocity $v$ and a massless antiquark with light-like momentum $p^\mu=\omega'\,\bn^\mu/2$, such that $n\cdot p=2v\cdot p=\omega'$. Without loss of generality, we choose $v^\mu=(1,0,0,0)$, $n^\mu=(1,0,0,1)$ and $\bn^\mu=2v^\mu-n^\mu$. Projecting onto the spinor structure of the tree-level matrix element, we define
\begin{equation}\label{eq:amp}
\langle 0|\,\cO_+^{\rm bare}(\omega)\,|b(v)\,\bar q(p)\rangle=\mathcal{A}(\omega,\omega',\epsilon)\;\bar v(p)\,\slashed{n}\gamma_5\,u(v)\,,
\end{equation}
with $\mathcal{A}^{(0)}(\omega,\omega',\epsilon)=i/(\omega'-\omega)$ at tree level. The corresponding matrix element of $O_+^{\rm bare}(\omega)$ is then given by 
\begin{align}
\cM(\omega,\omega',\epsilon)=-\frac{1}{2\pi}{\rm Disc}_\omega\,\mathcal{A}(\omega,\omega',\epsilon)\,. 
\end{align}

Following~\cite{Liu:2020ydl}, we define the radiative jet function in~\eqref{eq:Bgamma} in terms of the matrix element
\begin{align}\label{eq:Jdef}
\int d^dx&\,e^{i p_{s+}\cdot x}\langle\gamma(k)|\,T\big(\slashed{\mathscr{A}}_n^\perp(x)+\slashed{\mathscr{G}}_n^\perp(x)\big)\,\mathscr{X}_n(x)\,\bar{\mathscr{X}}_n(0)\,|0\rangle \nn\\
&= e\,\slashed{\varepsilon}^*_\perp(k)\,\frac{\slashed{n}}{2}\,\frac{i\,\bar n\cdot k}{p^2+i0}\,J(p^2)\,,
\end{align}
where $\mathscr{X}_n$, $\mathscr{A}_n$ and $\mathscr{G}_n$ are the gauge-invariant $n$-collinear quark, photon and gluon fields, respectively, with momentum scaling $(n\cdot p,\bn\cdot p,p_\perp)\sim(\omega, E_\gamma, \sqrt{E_\gamma\omega})$, and $e$ is the electromagnetic coupling. The momenta of the photon and the soft quark are denoted by $k$ and $p_s$, respectively. Here $n$ is along the photon momentum in $B^-\to\gamma\ell^-\bar\nu$, which coincides with the light-like vector of the LCDA in~\eqref{eq:LCDAdef}. The jet function depends only on $p^2\equiv(k+p_{s+})^2=\bar n\cdot k\,n\cdot p_s$ with $p_{s+}^\mu=(n\cdot p_s)\,\bar n^\mu/2$\,. As shown in \eqref{eq:Jdef}, the jet function can be decomposed into $J(p^2)=J_A(p^2)+J_G(p^2)$ at the bare level, where $J_A$ and $J_G$ correspond to the contributions from the photon and gluon fields, respectively, and they are gauge invariant individually.

\paragraph*{Method of IR subtraction.---}
For the on-shell partonic matrix element in dimensional regularization with $d=4-2\epsilon$, both UV and IR poles are present in the higher-order corrections to $\mathcal{A}(\omega,\omega',\epsilon)$ defined in~\eqref{eq:amp}. In the conventional approach, the IR poles can be regularized by assigning residual off-shell momenta to the external light antiquark and heavy quark, e.g.\ by $p^\mu\to p^\mu + \lambda\, n^\mu/2$ (such that $p^2=\omega'\lambda$) and $1/(v\cdot k)\to 1/(v\cdot k - \lambda/2)$ with $\lambda  \ll \omega, \omega'$. At leading power in $\lambda/\omega'$, the hard fluctuations with virtualities of order $\omega$ and $\omega'$ can be integrated out, and the operator $\cO_+^{\rm bare}(\omega)$ is matched onto a local heavy-light current in the low-energy effective theory,
\begin{align}\label{eq:matching}
\cO_+^{\rm bare}(\omega) &\to \mathcal{A}(\omega,\omega',\epsilon)\, \nn\\
&\quad\times\bar\xi_{\bn}(0)\,W_{\bn}(0) Y_{\bn}^\dagger(0) \slashed{n}
\gamma_5 Y_v(0) h_v^{(0)}(0)\,, 
\end{align}
where $\xi_{\bn}$ is the $\bn$-collinear quark field with momentum scaling $(\omega', \lambda, \sqrt{\omega'\lambda})$, carrying the total light-cone momentum $n\cdot p=\omega'$, $W_{\bn}(x)$ is the collinear Wilson line, and $Y_{\bn}(x)$ and $Y_v(x)$ are the soft Wilson lines along the $\bn$ and $v$ directions arising from the decoupling transformation~\cite{Bauer:2001yt}. $Y_{\bn}$ and $Y_v$ are defined in terms of the soft gluon field with momentum scaling $(\lambda, \lambda, \lambda)$, and should not be confused with $S_n$ in~\eqref{eq:LCDAdef}. The sterile field $h_v^{(0)}$ does not interact with soft gluons. The matching coefficient is the on-shell amplitude $\mathcal{A}(\omega,\omega',\epsilon)$, since the on-shell matrix elements in the low-energy theory are scaleless and thus equal to their tree-level values~\cite{Becher:2009qa}. Since the $\bn$-collinear fields are decoupled from the soft fields in~\eqref{eq:matching}, the IR-regularized matrix element takes the factorized form  
\begin{align}\label{eq:ampUV}
\langle 0|&\,\cO_+^{\rm bare}(\omega)\,|b(v)\,\bar q(p)\rangle\big|_{\rm UV}
\nn\\
&=\mathcal{A}(\omega,\omega',\epsilon)\,\cS(\lambda,\{v,\bn\})\,
\cJ(\omega'\lambda)\,{\bar v}(p)\,\slashed{n}\,\gamma_5\,u(v)\;,
\end{align}
where the jet function $\cJ(\omega'\lambda)$ is defined in terms of $\xi_{\bn}$ and $W_{\bn}$, and the soft function $\cS(\lambda,\{v,\bn\})=\langle 0|\,{\rm tr}\big[Y_{\bn}^\dagger(0)\,Y_v(0)\big]|0\rangle/N_c$ is the vacuum expectation value of the soft Wilson lines. Since the left-hand side of~\eqref{eq:ampUV} contains only UV poles, the UV poles of $\cS\,\cJ$ must cancel the IR poles of $\mathcal{A}(\omega,\omega',\epsilon)$.
Note that $\cJ(\omega'\lambda)$ should not be confused with the radiative jet function $J(p^2)$ in~\eqref{eq:Jdef}.

Let us now turn to the IR singularities of the heavy-light quark form factor. We consider the slightly off-shell amputated Green's function of the heavy-light current in QCD, which factorizes into the product of the on-shell heavy-light quark form factor, a jet function describing collinear emissions from the light quark, and a soft function describing soft-gluon exchange between the light and heavy quarks~\cite{Becher:2009qa,Becher:2009kw}. These jet and soft functions are identical to $\cJ$ and $\cS$ in~\eqref{eq:ampUV}, because the IR behavior of soft-gluon emissions from a heavy quark in QCD is identical to that in HQET. Consequently, the IR poles of $\mathcal{A}(\omega,\omega',\epsilon)$ are governed by the same anomalous dimensions as those of the heavy-light quark form factor. As shown in~\cite{Becher:2009qa,Becher:2009kw}, the UV poles of $\cJ$ and $\cS$ are governed by the collinear anomalous dimension $\Gamma_c(\ln(\omega'\lambda/\mu^2),\mu)$ and the soft anomalous dimension $\Gamma_s(\ln(\lambda/\mu),\mu)$, respectively. The IR poles of $\mathcal{A}(\omega,\omega',\epsilon)$ are therefore removed by the inverse of the IR renormalization factor $Z_{\rm hl}$, which can be derived from
\begin{align}\label{eq:ZhlRGE}
Z_{\rm hl}(\omega', \mu)=\exp \int_\mu^\infty \frac{d \bar \mu}{\bar \mu}\Gamma_{\rm hl}(\omega',\bar\mu)\,, 
\end{align}
with $\Gamma_{\rm hl}=\Gamma_c + \Gamma_s$. In the ${\overline {\rm MS}}$ scheme, $Z_{\rm hl}$ contains only poles in $\epsilon$. The $\lambda$ dependence cancels in $\Gamma_{\rm hl}$, the structure of which is given by~\cite{Becher:2009kw}
\begin{equation}\label{eq:Gammahl}
\Gamma_{\rm hl}(\omega',\mu)=-\Gamma_{\rm cusp}(\alpha_s)\ln\frac{\mu}{-\omega'-i0}+\gamma^q(\alpha_s)+\gamma^Q(\alpha_s)\,,
\end{equation}
where we have used $2v\cdot p=\omega'$, and the $-i0$ prescription accounts for the fact that both partons are incoming. The anomalous dimensions $\Gamma_{\rm cusp}$, $\gamma^q$ and $\gamma^Q$ have been determined to three-loop order~\cite{Moch:2004pa,Moch:2005id,Moch:2005tm,Becher:2006mr,Gehrmann:2010ue,Grozin:2014hna,Grozin:2015kna,Bruser:2019yjk}. 
Explicit expressions for $\ln Z_{\rm hl}$ and for these anomalous dimensions up to three-loop order are collected in the Supplemental Material.

Since $Z_{\rm hl}$ depends only on $\omega'$, it commutes both with the discontinuity in $\omega$ and with the convolution in~\eqref{eq:Zplus}. The renormalization factor of $O_+^{\rm bare}(\omega)$ is then determined by the condition
\begin{equation}\label{eq:finite}
\int d\omega''\,Z_+(\omega,\omega'',\mu)\,\cM(\omega'',\omega',\epsilon)\,Z_{\rm hl}^{-1}(\omega',\mu) 
= \mathcal{O}(\epsilon^0)\,.
\end{equation}
In this approach, the UV poles are extracted from the on-shell matrix elements of a gauge-invariant operator, without any additional IR regulator. 
The remaining task is to calculate the on-shell partonic matrix element $\cM(\omega,\omega',\epsilon)$ up to three loops. Together with $Z_{\rm hl}$, which is already known to $\mathcal{O}(\alpha_s^3)$, the renormalization factor $Z_+$ and the evolution kernel $\gamma_+$ can then be determined order by order up to three loops. We have successfully reproduced the one- and two-loop evolution kernels~\cite{Lange:2003ff,Braun:2019wyx}, which demonstrates the validity of our method of IR subtraction.

\paragraph*{Computational method.---}
Based on the partonic matrix element defined in~\eqref{eq:amp}, we generate the Feynman diagrams with \texttt{QGRAF}~\cite{Nogueira:1991ex}. There are 702 diagrams contributing at three loops. The diagrams are further processed by an in-house \texttt{Mathematica} code, which assigns the corresponding Feynman rules and performs the Dirac, Lorentz and color algebra. After partial-fraction decompositions, all scalar Feynman integrals are mapped onto twenty integral topologies, each consisting of nine quadratic and six linear propagators, three of the latter along $v$ and three along $n$. We perform the IBP reduction~\cite{Tkachov:1981wb,Chetyrkin:1981qh} for the integrals in each of
the twenty topologies by using the public programs \texttt{FIRE6}~\cite{Smirnov:2019qkx} and \texttt{Kira}~\cite{Klappert:2020nbg}. Besides the eighteen IBP identities in each topology, obtained by inserting differential operators in front of the integrands, three more linear algebraic identities following from the relation between the external vectors, $p^\mu-\omega'(v^\mu-n^\mu/2)=0$, have been implemented into the reduction. After identifying equivalent master integrals (MIs) in different topologies, the three-loop matrix element is finally expressed as a linear combination of 556 MIs.

In the next step, we use the canonical DE method~\cite{Kotikov:1990kg,Remiddi:1997ny,Gehrmann:1999as,Henn:2013pwa} to solve for the MIs. To accelerate the search for a canonical basis, we first perform a leading-singularity analysis in the Baikov representation to identify uniform transcendental (UT) integral candidates for selected sectors~\cite{Baikov:1996rk,Arkani-Hamed:2010pyv}, and replace the original MIs with these UT candidates.
By using the public packages \texttt{CANONICA}~\cite{Meyer:2017joq} and \texttt{Libra}~\cite{Lee:2020zfb}, the DE system of the 556 MIs is then converted into a canonical form 
\begin{equation}\label{eq:canonical}
d\vec I(\omega,\omega';\epsilon)=\epsilon\,\left(\sum_{k=1}^{14}c_k\,d\log W_k(\omega,\omega')\right)\vec I(\omega,\omega';\epsilon)\,,
\end{equation}
where the $c_k$ are constant matrices and the $W_k$ are the corresponding letters. Fourteen letters appear in the full DE system, while only four of them are present in the three-loop ${\cal A}(\omega,\omega',\epsilon)$ expanded up to $\mathcal{O}(\epsilon^{-1})$, which are given by $\{\omega, \omega', \omega-\omega', (\sqrt{\omega'}+\sqrt{\omega'-\omega})/(\sqrt{\omega'}-\sqrt{\omega'-\omega})\}$.
By using the method of regions~\cite{Beneke:1997zp}, we determine the boundary conditions in the limit $\omega'\ll\omega$, in which each loop momentum is either hard, $k_i\sim\omega$, or soft, $k_i\sim\omega'$. 
The integrals in each region depend on a single scale, so that their scale dependence is trivial. By performing IBP reduction and building dimensional recurrence relations (DRRs)~\cite{Tarasov:1996br,Lee:2009dh,Lee:2012cn}, the regional integrals are mapped onto quasi-finite integrals in $d=6-2\epsilon$ or $d=8-2\epsilon$ dimensions~\cite{vonManteuffel:2014qoa}, which can be evaluated analytically by integrating over Feynman parameters with \texttt{HyperInt}~\cite{Panzer:2014caa}. With these boundary conditions and the canonical DEs in~\eqref{eq:canonical}, we then solve for the MIs order by order in $\epsilon$ in terms of Goncharov polylogarithms (GPLs)~\cite{Goncharov:1998kja} and generalized harmonic polylogarithms (GHPLs)~\cite{Aglietti:2004tq}, up to transcendental weight five. 

Inserting the MIs, we obtain the bare three-loop matrix element $\mathcal{A}(\omega,\omega',\epsilon)$ up to $\mathcal{O}(\epsilon^{-1})$ for all color structures. The one-loop and two-loop matrix elements are also obtained up to $\mathcal{O}(\epsilon^3)$ and $\mathcal{O}(\epsilon)$, respectively. After the renormalization of $\alpha_s$ and the subtraction of the IR singularities by $Z_{\rm hl}^{-1}$, we take the discontinuity in $\omega$. The convolution in~\eqref{eq:finite} is performed with the help of \texttt{PolyLogTools}~\cite{Duhr:2019tlz}. The discontinuities across the branch cuts on the positive real $\omega$ axis, which start at the branch points $\omega=0$ and $\omega=\omega'$, determine the non-local terms of the evolution kernel, while the endpoint $\omega=\omega'$ gives rise to the $\delta$-function and plus-distribution terms. Finally, at three-loop order, the renormalization factor $Z_+$ is determined from the condition~\eqref{eq:finite}, and the evolution kernel is then derived from~\eqref{eq:gamma}.

Let us briefly discuss the calculation of the radiative jet function defined in~\eqref{eq:Jdef}. Since soft radiation has already been decoupled from the collinear SCET Lagrangian by means of a field redefinition, we can use standard QCD Feynman rules. By using \texttt{QGRAF}, we generate 3348 three-loop diagrams in Feynman gauge. The procedure of the amplitude computation is very similar to that in~\cite{Bruser:2018rad,Liu:2020ydl}. After performing IBP reduction and using the DRRs, the three-loop jet function is expressed in terms of 69 quasi-finite integrals, which are found by the public package \texttt{Reduze}~\cite{vonManteuffel:2012np} and then evaluated analytically with \texttt{HyperInt}. All MIs are then checked numerically with \texttt{AMFlow}~\cite{Liu:2017jxz,Liu:2022chg}. 
Finally, we obtain the bare radiative jet function expanded to $\mathcal{O}(\epsilon^0)$ at three loops, the result of which is presented in the Supplemental Material.

\paragraph*{Results and consistency checks.---}
Combining the one- and two-loop matrix elements with the three-loop cusp anomalous dimension, the two-loop evolution kernel and the three-loop $Z_{\rm hl}$, we predict the poles from $1/\epsilon^6$ to $1/\epsilon^2$ of the IR-subtracted three-loop matrix element $\cM\,Z_{\rm hl}^{-1}$, which agree completely with our three-loop calculation. This provides a highly non-trivial check of both the three-loop calculation and the method of IR subtraction. The remaining $1/\epsilon$ poles then determine, for the first time, the three-loop evolution kernel of the $B$-meson LCDA. Up to three-loop order, the evolution kernel in~\eqref{eq:RGE} takes the form
\begin{align}\label{eq:gamma3}
\gamma_+(\omega,\omega',\mu)&=\Big[\Gamma_{\rm cusp}(\alpha_s)\ln\frac{\mu}{\omega}+\gamma(\alpha_s)\Big]\delta(\omega-\omega') \nn\\
&+\omega\,\Gamma_{\rm cusp}(\alpha_s)\,\Gamma(\omega,\omega')\nn\\
&+\sum_{n=2,3}\left(\frac{\alpha_s}{4\pi}\right)^n 
\frac{\omega\,\theta(\omega'-\omega)}{\omega'(\omega'-\omega)}
{h}^{(n)}\left(\frac{\omega}{\omega'}\right)
\,,
\end{align}
where
\begin{equation}\label{eq:LNkernel}
\Gamma(\omega,\omega')=-\left[\frac{\theta(\omega-\omega')}{\omega(\omega-\omega')}+\frac{\theta(\omega'-\omega)}{\omega'(\omega'-\omega)}\right]_+\,.
\end{equation}
The plus distribution implies the replacement $f(\omega')\to f(\omega')-f(\omega)$ when the kernel is integrated with a function $f(\omega')$~\cite{Lange:2003ff}. 
The coefficients of the non-cusp anomalous dimension $\gamma$, as well as the functions $h^{(n)}(x)$ with $x=\omega/\omega'\in(0,1)$, are given up to three loops in the Supplemental Material. 

The three-loop extension of the Lange-Neubert kernel given in~\eqref{eq:gamma3} exhibits several remarkable features. 
First, we find that the coefficient of the logarithm $\ln(\mu/\omega)$ coincides with the light-like cusp anomalous dimension up to three-loop order~\cite{Korchemsky:1987wg,Moch:2004pa}, as required by the RG invariance of the convolution in~\eqref{eq:conv}. 
In addition, $h^{(n)}(\omega/\omega')$ vanishes for $\omega\to\omega'$, so that the plus prescription is not needed for the non-local terms corresponding to $h^{(n)}(\omega/\omega')$ in $\gamma_+$. Our two-loop coefficient of the non-cusp anomalous dimension $\gamma$ and the function $h^{(2)}$ are consistent with the two-loop evolution kernel in~\cite{Braun:2019wyx}. Finally, although the bare three-loop matrix element $\mathcal{A}(\omega,\omega',\epsilon)$ contains GHPLs with the square-root letter, they cancel out when combined with the $\mathcal{O}(\epsilon)$ terms of the two-loop $\mathcal{A}(\omega,\omega',\epsilon)$ during renormalization. After taking the discontinuity, the three-loop function $h^{(3)}(x)$ is expressed in terms of logarithms and the classical polylogarithms $\mathrm{Li}_2$ and $\mathrm{Li}_3$ only. We emphasize that the local and non-local contributions to the kernel are fully determined by taking the discontinuity. 
The analytic result of ${\cal A}(\omega,\omega',\epsilon)$ and the procedure for taking the discontinuity are made publicly available in~\cite{zenodo}.

In order to demonstrate the validity of the evolution kernel, we check its consistency with the renormalization of the radiative jet function up to three loops. Using the RG invariance of the decay amplitude and working with the same convention as in~\cite{Liu:2020ydl}, we obtain the anomalous dimension $\gamma_J$ of the radiative jet function at three loops, which then determines the renormalization factor $Z_J$ of the jet function.   
At the same time, we calculate the three-loop bare radiative jet function $J^{(0)}(p^2)$ independently. 
We find that $Z_J$ removes all poles of the three-loop bare radiative jet function, from $1/\epsilon^6$ to $1/\epsilon$ exactly. This not only demonstrates the validity of our result for the three-loop Lange-Neubert kernel given in \eqref{eq:gamma3}, but also establishes the RG consistency of the factorization theorem for the leptonic radiative $B$ decay, because the $1/\epsilon$ poles of $Z_J$ are directly sensitive to $\gamma$ and $h^{(3)}(x)$.
It was observed in~\cite{Liu:2020ydl} that $\gamma'=-\gamma_W$ holds up to two loops, where $\gamma'$ and $\gamma_W$ are the non-cusp anomalous dimensions of the radiative jet function and of the soft function for Drell-Yan production near threshold, respectively. Interestingly, we find that this relation persists at three loops~\cite{Moch:2004pa,Becher:2007ty}.         
The expressions of $\gamma_J$, $\gamma'$ and $J^{(0)}(p^2)$ are presented in the Supplemental Material.

\paragraph*{Phenomenological impact.---}
\begin{figure}[t]
\begin{center}
\includegraphics[width=\linewidth]{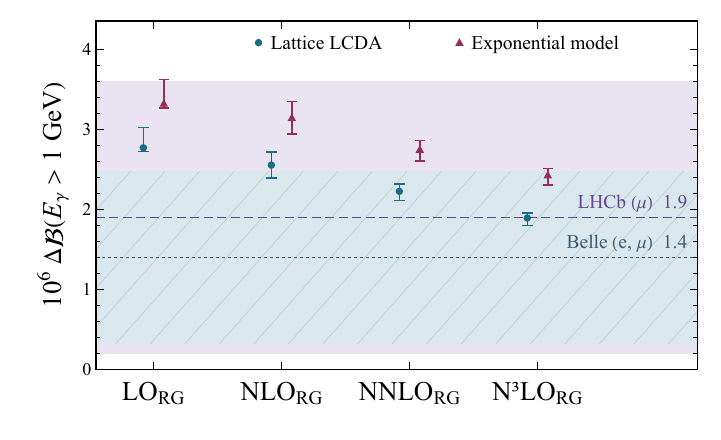}
\end{center}
\captionsetup{justification=raggedright}
\caption{\label{fig:phenobr}
Partial branching fraction of $B^-\to\gamma\ell^-\bar\nu_\ell$ with $E_\gamma>1\,\mathrm{GeV}$. The error bars represent the scale uncertainties obtained by varying $\mu_h$ from $m_b^{\rm pole}/2$ to $2\,m_b^{\rm pole}$.
The green dotted and purple dashed lines with the bands represent the measurements from Belle~\cite{Belle:2018jqd} and LHCb~\cite{LHCb:2026clt}, respectively.  
}
\end{figure}

To illustrate the phenomenological impact of the three-loop radiative jet function, following~\cite{Liu:2020ydl}, we evaluate its convolution with the $B$-meson LCDA, 
\begin{equation}
\frac{J\otimes\phi}{e_q}\simeq
\frac{1}{\lambda_B(\mu_j)}
\left( 1 + 3.94\%|_{\alpha_s} -  1.65\%|_{\alpha_s^2} -  4.22\%|_{\alpha_s^3}
\right)
\,,
\label{eq:jetphen3lnum}
\end{equation}
with the input parameters chosen as $\mu=\mu_j=1.87\,\mathrm{GeV}$, $E_\gamma=1.75\,\mathrm{GeV}$ and $n_f=4$. We have adopted the logarithmic moment expansion of the $B$-meson LCDA with the inverse moment $\lambda_B(\mu_j)$~\cite{Beneke:2011nf}. The non-perturbative input for the LCDA is obtained from lattice QCD calculations~\cite{LPC:2026vyv}. In the Supplemental Material, we present the details on the phenomenological impact of the radiative jet function, together with its RG evolution at three loops. 

We also compare the RG-improved predictions for the partial branching fraction of $B^-\to\gamma\ell^-\bar\nu_\ell$ with the experimental measurements from Belle~\cite{Belle:2018jqd} and LHCb~\cite{LHCb:2026clt}. 
We adopt two different non-perturbative inputs for the $B$-meson LCDA: one is from the lattice $K=5$ Laguerre reconstruction at $1.87\,\mathrm{GeV}$~\cite{LPC:2026vyv}, and the other is from the Grozin--Neubert exponential model~\cite{Grozin:1996pq} with $\lambda_B(\mu_j)$ evolved from $\lambda_B(1\,\mathrm{GeV})=0.350\,\mathrm{GeV}$.
The hard matching coefficient in \eqref{eq:Bgamma} is evolved from the hard scale $\mu_h$ to the jet scale $\mu_j$ with four-loop $\Gamma_{\rm cusp}$~\cite{Henn:2019swt,vonManteuffel:2020vjv} and three-loop $\gamma_F$, $\gamma^q$ and $\gamma^Q$~\cite{Chetyrkin:2003vi,Moch:2005id,Grozin:2014hna,Grozin:2015kna,Bruser:2019yjk}. Combining these with the three-loop radiative jet function and heavy-light form factor~\cite{Fael:2024vko}, we obtain the RG-improved N$^3$LO predictions for the partial branching fraction. 
As shown in Fig.~\ref{fig:phenobr}, the higher-order corrections decrease the partial branching fraction for both LCDA inputs, and the RG-improved N$^3$LO predictions are compatible with the Belle and LHCb measurements within their experimental uncertainties. The uncertainties from varying $\mu_h$ are reduced at higher orders. The significant N$^3$LO corrections imply the importance of the three-loop radiative jet function in the precise determination of $\lambda_B$. The input parameters are collected in Supplemental Material.

\paragraph*{Conclusion.---}
In this {\it Letter}, we have presented the calculation of the three-loop evolution kernel of the $B$-meson LCDA. The $B$-meson LCDA in momentum space is represented as the discontinuity of a matrix element with an inverse differential operator. The corresponding on-shell partonic matrix element is expressed in terms of standard Feynman integrals, which enables us to perform the calculation by means of modern techniques for multi-loop Feynman integrals. We have shown that the IR singularities of this matrix element are governed by the same anomalous dimension as that of the heavy-light quark form factor. 
Therefore, they are removed by the known renormalization factor $Z_{\rm hl}$, without introducing any IR regulator that would violate gauge invariance or complicate the Feynman integrals. In principle, the novel approach to IR subtraction can be applied to other non-local operators, such as those defining QED-generalized and higher-twist LCDAs of $B$ mesons~\cite{Braun:2017liq,Ferre:2026sao}.

The three-loop radiative jet function gives sizable corrections to the $B^-\to\gamma\ell^-\bar\nu$ decay amplitude. The RG-improved N$^3$LO prediction reduces perturbative uncertainties in determinations of the non-perturbative $B$-meson LCDA. The three-loop RG evolution of the radiative jet function also helps to deepen our understanding of renormalization in subleading-power factorization theorems~\cite{Beneke:2017ztn,Moult:2019mog,Liu:2019oav,Liu:2020wbn,Liu:2020tzd,Laenen:2020nrt,Beneke:2022obx,Beneke:2024cpq}.

\vspace{1mm}
\paragraph*{Acknowledgments.---}
We are grateful to Long Chen for helpful discussions. This work is supported by Institute of High Energy Physics (IHEP, CAS) under Grant No. E35159U1. We used OpenAI GPT-6 Astra to assist in working with \texttt{Libra}, and Claude Opus 5.5 in an editorial capacity.

\vspace{1mm}
\paragraph*{Note added.---}
While finalizing this manuscript, we became aware of independent calculations of the three-loop evolution kernel and the radiative jet function using different methods~\cite{Huang:2026fck,Ji:2026jwu}, which appear on the arXiv simultaneously with this {\it Letter}.

\bibliographystyle{apsrev4-1}
\bibliography{letter}

@article{Becher:2007ty,
      author         = "Becher, Thomas and Neubert, Matthias and Xu, Gang",
      title          = "{Dynamical Threshold Enhancement and Resummation in
                        Drell-Yan Production}",
      journal        = "JHEP",
      volume         = "07",
      year           = "2008",
      pages          = "030",
      doi            = "10.1088/1126-6708/2008/07/030",
      eprint         = "0710.0680",
      archivePrefix  = "arXiv",
      primaryClass   = "hep-ph",
      reportNumber   = "CLNS-07-2009, FERMILAB-PUB-07-507-T, MZ-TH-07-16",
      SLACcitation   = "%%CITATION = ARXIV:0710.0680;%%"
}

@article{vonManteuffel:2012np,
    author = "von Manteuffel, A. and Studerus, C.",
    title = "{Reduze 2 - Distributed Feynman Integral Reduction}",
    eprint = "1201.4330",
    archivePrefix = "arXiv",
    primaryClass = "hep-ph",
    reportNumber = "ZU-TH-01-12, MZ-TH-12-03, BI-TP-2012-02",
    month = "1",
    year = "2012"
}

@article{Lee:2012cn,
    author = "Lee, R. N.",
    title = "{Presenting LiteRed: a tool for the Loop InTEgrals REDuction}",
    eprint = "1212.2685",
    archivePrefix = "arXiv",
    primaryClass = "hep-ph",
    month = "12",
    year = "2012"
}

@article{Henn:2019swt,
    author = "Henn, Johannes M. and Korchemsky, Gregory P. and Mistlberger, Bernhard",
    title = "{The full four-loop cusp anomalous dimension in $\mathcal{N}=4$ super Yang-Mills and QCD}",
    eprint = "1911.10174",
    archivePrefix = "arXiv",
    primaryClass = "hep-th",
    reportNumber = "IPhT-T19/155, MIT-CTP/5160, MPP-2019-233",
    doi = "10.1007/JHEP04(2020)018",
    journal = "JHEP",
    volume = "04",
    pages = "018",
    year = "2020"
}

@article{vonManteuffel:2020vjv,
    author = "von Manteuffel, Andreas and Panzer, Erik and Schabinger, Robert M.",
    title = "{Cusp and collinear anomalous dimensions in four-loop QCD from form factors}",
    eprint = "2002.04617",
    archivePrefix = "arXiv",
    primaryClass = "hep-ph",
    reportNumber = "MSUHEP-20-002",
    doi = "10.1103/PhysRevLett.124.162001",
    journal = "Phys. Rev. Lett.",
    volume = "124",
    number = "16",
    pages = "162001",
    year = "2020"
}

@article{Moch:2005tm,
    author = "Moch, S. and Vermaseren, J. A. M. and Vogt, A.",
    title = "{Three-loop results for quark and gluon form-factors}",
    eprint = "hep-ph/0508055",
    archivePrefix = "arXiv",
    reportNumber = "DESY-05-138, SFT-CPP-05-34, NIKHEF-05-013, DCPT-05-88, IPPP-05-44",
    doi = "10.1016/j.physletb.2005.08.067",
    journal = "Phys. Lett. B",
    volume = "625",
    pages = "245--252",
    year = "2005"
}

@article{Aglietti:2004tq,
    author = "Aglietti, U. and Bonciani, R.",
    title = "{Master integrals with 2 and 3 massive propagators for the 2 loop electroweak form-factor - planar case}",
    eprint = "hep-ph/0401193",
    archivePrefix = "arXiv",
    reportNumber = "ROME1-1367-03, FREIBURG-THEP-04-02",
    doi = "10.1016/j.nuclphysb.2004.07.018",
    journal = "Nucl. Phys. B",
    volume = "698",
    pages = "277--318",
    year = "2004"
}

@article{Tarasov:1996br,
    author = "Tarasov, O. V.",
    title = "{Connection between Feynman integrals having different values of the space-time dimension}",
    eprint = "hep-th/9606018",
    archivePrefix = "arXiv",
    reportNumber = "DESY-96-068, JINR-E2-96-62",
    doi = "10.1103/PhysRevD.54.6479",
    journal = "Phys. Rev. D",
    volume = "54",
    pages = "6479--6490",
    year = "1996"
}

@article{Lee:2009dh,
    author = "Lee, R. N.",
    title = "{Space-time dimensionality D as complex variable: Calculating loop integrals using dimensional recurrence relation and analytical properties with respect to D}",
    eprint = "0911.0252",
    archivePrefix = "arXiv",
    primaryClass = "hep-ph",
    doi = "10.1016/j.nuclphysb.2009.12.025",
    journal = "Nucl. Phys. B",
    volume = "830",
    pages = "474--492",
    year = "2010"
}

@article{Liu:2017jxz,
    author = "Liu, Xiao and Ma, Yan-Qing and Wang, Chen-Yu",
    title = "{A Systematic and Efficient Method to Compute Multi-loop Master Integrals}",
    eprint = "1711.09572",
    archivePrefix = "arXiv",
    primaryClass = "hep-ph",
    doi = "10.1016/j.physletb.2018.02.026",
    journal = "Phys. Lett. B",
    volume = "779",
    pages = "353--357",
    year = "2018"
}

@article{Liu:2022chg,
    author = "Liu, Xiao and Ma, Yan-Qing",
    title = "{AMFlow: A Mathematica package for Feynman integrals computation via auxiliary mass flow}",
    eprint = "2201.11669",
    archivePrefix = "arXiv",
    primaryClass = "hep-ph",
    doi = "10.1016/j.cpc.2022.108565",
    journal = "Comput. Phys. Commun.",
    volume = "283",
    pages = "108565",
    year = "2023"
}

@misc{zenodo,
  author       = {Liu, Ze Long and Tang, Guang and Wang, Zhe-Yu},
  title        = {Ancillary files provided with the submission of this article},
  month        = nov,
  year         = 2026,
  publisher    = {Zenodo},
  note         = {https://doi.org/10.5281/zenodo.24468147},
  doi          = {10.5281/zenodo.24468147},
  url          = {https://doi.org/10.5281/zenodo.24468147},
}

@misc{amflow,
  author = {},
  title = {},
  year = {},
  howpublished = {\url{https://gitlab.com/multiloop-pku/amflow}},
  note = {}
}

@article{Bruser:2018rad,
	author         = "Brüser, Robin and Liu, Ze Long and Stahlhofen,
	Maximilian",
	title          = "{Three-Loop Quark Jet Function}",
	journal        = "Phys. Rev. Lett.",
	volume         = "121",
	year           = "2018",
	number         = "7",
	pages          = "072003",
	doi            = "10.1103/PhysRevLett.121.072003",
	eprint         = "1804.09722",
	archivePrefix  = "arXiv",
	primaryClass   = "hep-ph",
	reportNumber   = "MITP/18-031, MITP-18-031",
	SLACcitation   = "%%CITATION = ARXIV:1804.09722;%%"
}

@article{Duhr:2019tlz,
    author = "Duhr, Claude and Dulat, Falko",
    title = "{PolyLogTools {\textemdash} polylogs for the masses}",
    eprint = "1904.07279",
    archivePrefix = "arXiv",
    primaryClass = "hep-th",
    reportNumber = "CP3-19-17, CERN-TH-2019-045, SLAC-PUB-17423",
    doi = "10.1007/JHEP08(2019)135",
    journal = "JHEP",
    volume = "08",
    pages = "135",
    year = "2019"
}

@article{Korchemsky:1987wg,
    author = "Korchemsky, G. P. and Radyushkin, A. V.",
    title = "{Renormalization of the Wilson Loops Beyond the Leading Order}",
    doi = "10.1016/0550-3213(87)90277-X",
    journal = "Nucl. Phys. B",
    volume = "283",
    pages = "342--364",
    year = "1987"
}

@article{Chetyrkin:1981qh,
    author = "Chetyrkin, K. G. and Tkachov, F. V.",
    title = "{Integration by parts: The algorithm to calculate $\beta$-functions in 4 loops}",
    doi = "10.1016/0550-3213(81)90199-1",
    journal = "Nucl. Phys. B",
    volume = "192",
    pages = "159--204",
    year = "1981"
}

@article{Kotikov:1990kg,
    author = "Kotikov, A. V.",
    title = "{Differential equations method: New technique for massive Feynman diagrams calculation}",
    reportNumber = "ITF-90-31E",
    doi = "10.1016/0370-2693(91)90413-K",
    journal = "Phys. Lett. B",
    volume = "254",
    pages = "158--164",
    year = "1991"
}

@article{Tkachov:1981wb,
    author = "Tkachov, F. V.",
    title = "{A theorem on analytical calculability of 4-loop renormalization group functions}",
    doi = "10.1016/0370-2693(81)90288-4",
    journal = "Phys. Lett. B",
    volume = "100",
    pages = "65--68",
    year = "1981"
}

@article{Broadhurst:1991fz,
    author = "Broadhurst, David J. and Grozin, A. G.",
    title = "{Two loop renormalization of the effective field theory of a static quark}",
    eprint = "hep-ph/9908362",
    archivePrefix = "arXiv",
    reportNumber = "OUT-4102-30",
    doi = "10.1016/0370-2693(91)90532-U",
    journal = "Phys. Lett. B",
    volume = "267",
    pages = "105--110",
    year = "1991"
}

@article{Nogueira:1991ex,
    author = "Nogueira, Paulo",
    title = "{Automatic Feynman Graph Generation}",
    reportNumber = "IFM-7-91",
    doi = "10.1006/jcph.1993.1074",
    journal = "J. Comput. Phys.",
    volume = "105",
    pages = "279--289",
    year = "1993"
}

@article{Bosch:2003fc,
    author = "Bosch, S. W. and Hill, R. J. and Lange, B. O. and Neubert, M.",
    title = "{Factorization and Sudakov resummation in leptonic radiative B decay}",
    eprint = "hep-ph/0301123",
    archivePrefix = "arXiv",
    reportNumber = "SLAC-PUB-9625, CLNS-03-1817",
    doi = "10.1103/PhysRevD.67.094014",
    journal = "Phys. Rev. D",
    volume = "67",
    pages = "094014",
    year = "2003"
}

@article{Smirnov:2019qkx,
    author = "Smirnov, A. V. and Chukharev, F. S.",
    title = "{FIRE6: Feynman Integral REduction with modular arithmetic}",
    eprint = "1901.07808",
    archivePrefix = "arXiv",
    primaryClass = "hep-ph",
    reportNumber = "TTP19-006",
    doi = "10.1016/j.cpc.2019.106877",
    journal = "Comput. Phys. Commun.",
    volume = "247 ",
    pages = "106877",
    year = "2020"
}

@article{Bell:2013tfa,
    author = "Bell, Guido and Feldmann, Thorsten and Wang, Yu-Ming and Yip, Matthew W Y",
    title = "{Light-Cone Distribution Amplitudes for Heavy-Quark Hadrons}",
    eprint = "1308.6114",
    archivePrefix = "arXiv",
    primaryClass = "hep-ph",
    reportNumber = "OUTP-13-15P, SI-HEP-2013-05, QFET-2013-05, TTK-13-19, SFB-CPP-13-60, IPPP-13-55, DCPT-13-110, TUM-HEP-899-13",
    doi = "10.1007/JHEP11(2013)191",
    journal = "JHEP",
    volume = "11",
    pages = "191",
    year = "2013"
}

@article{Liu:2020ydl,
    author = "Liu, Ze Long and Neubert, Matthias",
    title = "{Two-Loop Radiative Jet Function for Exclusive $B$-Meson and Higgs Decays}",
    eprint = "2003.03393",
    archivePrefix = "arXiv",
    primaryClass = "hep-ph",
    reportNumber = "MITP/20-011",
    doi = "10.1007/JHEP06(2020)060",
    journal = "JHEP",
    volume = "06",
    pages = "060",
    year = "2020"
}

@article{Beneke:1997zp,
    author = "Beneke, M. and Smirnov, Vladimir A.",
    title = "{Asymptotic expansion of Feynman integrals near threshold}",
    eprint = "hep-ph/9711391",
    archivePrefix = "arXiv",
    reportNumber = "CERN-TH-97-315",
    doi = "10.1016/S0550-3213(98)00138-2",
    journal = "Nucl. Phys. B",
    volume = "522",
    pages = "321--344",
    year = "1998"
}

@article{Klappert:2020nbg,
    author = {Klappert, Jonas and Lange, Fabian and Maierh{\"o}fer, Philipp and Usovitsch, Johann},
    title = "{Integral reduction with Kira 2.0 and finite field methods}",
    eprint = "2008.06494",
    archivePrefix = "arXiv",
    primaryClass = "hep-ph",
    reportNumber = "TTK-20-24, P3H-20-041, FR-PHENO-2020-11, MITP/20-044",
    doi = "10.1016/j.cpc.2021.108024",
    journal = "Comput. Phys. Commun.",
    volume = "266",
    pages = "108024",
    year = "2021"
}

@article{Lee:2020zfb,
    author = "Lee, Roman N.",
    title = "{Libra: A package for transformation of differential systems for multiloop integrals}",
    eprint = "2012.00279",
    archivePrefix = "arXiv",
    primaryClass = "hep-ph",
    doi = "10.1016/j.cpc.2021.108058",
    journal = "Comput. Phys. Commun.",
    volume = "267",
    pages = "108058",
    year = "2021"
}

@article{Remiddi:1997ny,
    author = "Remiddi, Ettore",
    title = "{Differential equations for Feynman graph amplitudes}",
    eprint = "hep-th/9711188",
    archivePrefix = "arXiv",
    reportNumber = "DFUB-97-15, DFUB 97-15",
    doi = "10.1007/BF03185566",
    journal = "Nuovo Cim. A",
    volume = "110",
    pages = "1435--1452",
    year = "1997"
}

@article{Beneke:2023nmj,
    author = "Beneke, Martin and Finauri, Gael and Vos, K. Keri and Wei, Yanbing",
    title = "{QCD light-cone distribution amplitudes of heavy mesons from boosted HQET}",
    eprint = "2305.06401",
    archivePrefix = "arXiv",
    primaryClass = "hep-ph",
    reportNumber = "TUM-HEP-1455/23, Nikhef-2023-003",
    doi = "10.1007/JHEP09(2023)066",
    journal = "JHEP",
    volume = "09",
    pages = "066",
    year = "2023"
}

@article{Wang:2018wfj,
    author = "Wang, Yu-Ming and Shen, Yue-Long",
    title = "{Subleading-power corrections to the radiative leptonic $B \to \gamma \ell \nu$ decay in QCD}",
    eprint = "1803.06667",
    archivePrefix = "arXiv",
    primaryClass = "hep-ph",
    doi = "10.1007/JHEP05(2018)184",
    journal = "JHEP",
    volume = "05",
    pages = "184",
    year = "2018"
}

@article{Anastasiou:2002yz,
    author = "Anastasiou, Charalampos and Melnikov, Kirill",
    title = "{Higgs boson production at hadron colliders in NNLO QCD}",
    eprint = "hep-ph/0207004",
    archivePrefix = "arXiv",
    reportNumber = "SLAC-PUB-9273",
    doi = "10.1016/S0550-3213(02)00837-4",
    journal = "Nucl. Phys. B",
    volume = "646",
    pages = "220--256",
    year = "2002"
}

@article{Galda:2020epp,
    author = "Galda, Anne Mareike and Neubert, Matthias",
    title = "{Evolution of the $B$-Meson Light-Cone Distribution Amplitude in Laplace Space}",
    eprint = "2006.05428",
    archivePrefix = "arXiv",
    primaryClass = "hep-ph",
    reportNumber = "MITP/20-031",
    doi = "10.1103/PhysRevD.102.071501",
    journal = "Phys. Rev. D",
    volume = "102",
    pages = "071501",
    year = "2020"
}

@article{Galda:2022dhp,
    author = "Galda, Anne Mareike and Neubert, Matthias and Wang, Xing",
    title = "{Factorization and Sudakov resummation in leptonic radiative B decay {\textemdash} a reappraisal}",
    eprint = "2203.08202",
    archivePrefix = "arXiv",
    primaryClass = "hep-ph",
    reportNumber = "MITP/22-024",
    doi = "10.1007/JHEP07(2022)148",
    journal = "JHEP",
    volume = "07",
    pages = "148",
    year = "2022"
}

@article{Grozin:1996pq,
    author = "Grozin, A. G. and Neubert, M.",
    title = "{Asymptotics of heavy meson form-factors}",
    eprint = "hep-ph/9607366",
    archivePrefix = "arXiv",
    reportNumber = "BUDKER-INP-1996-45, BINP-96-45, CERN-TH-96-144",
    doi = "10.1103/PhysRevD.55.272",
    journal = "Phys. Rev. D",
    volume = "55",
    pages = "272--290",
    year = "1997"
}

@article{Belle:2018jqd,
    author = "Gelb, M. and others",
    collaboration = "Belle",
    title = "{Search for the rare decay of $B^+ \to \ell^{\,+} \nu_{\ell} \gamma$ with improved hadronic tagging}",
    eprint = "1810.12976",
    archivePrefix = "arXiv",
    primaryClass = "hep-ex",
    doi = "10.1103/PhysRevD.98.112016",
    journal = "Phys. Rev. D",
    volume = "98",
    number = "11",
    pages = "112016",
    year = "2018"
}

@article{Descotes-Genon:2002crx,
    author = "Descotes-Genon, S. and Sachrajda, C. T.",
    title = "{Factorization, the light cone distribution amplitude of the B meson and the radiative decay B ---{\ensuremath{>}} gamma l nu(l)}",
    eprint = "hep-ph/0209216",
    archivePrefix = "arXiv",
    reportNumber = "CERN-TH-2002-228, SHEP-02-16",
    doi = "10.1016/S0550-3213(02)01066-0",
    journal = "Nucl. Phys. B",
    volume = "650",
    pages = "356--390",
    year = "2003"
}

@article{Becher:2006mr,
    author = "Becher, Thomas and Neubert, Matthias and Pecjak, Ben D.",
    title = "{Factorization and Momentum-Space Resummation in Deep-Inelastic Scattering}",
    eprint = "hep-ph/0607228",
    archivePrefix = "arXiv",
    reportNumber = "CLNS-06-1971, FERMILAB-PUB-06-242-T, DFB-CPP-06-32, SI-HEP-2006-09",
    doi = "10.1088/1126-6708/2007/01/076",
    journal = "JHEP",
    volume = "01",
    pages = "076",
    year = "2007"
}

@article{Beneke:1999br,
    author = "Beneke, M. and Buchalla, G. and Neubert, M. and Sachrajda, Christopher T.",
    title = "{QCD factorization for B ---{\ensuremath{>}} pi pi decays: Strong phases and CP violation in the heavy quark limit}",
    eprint = "hep-ph/9905312",
    archivePrefix = "arXiv",
    reportNumber = "SLAC-PUB-8146, CERN-TH-99-126, SHEP-99-04",
    doi = "10.1103/PhysRevLett.83.1914",
    journal = "Phys. Rev. Lett.",
    volume = "83",
    pages = "1914--1917",
    year = "1999"
}

@article{Goncharov:1998kja,
    author = "Goncharov, Alexander B.",
    title = "{Multiple polylogarithms, cyclotomy and modular complexes}",
    eprint = "1105.2076",
    archivePrefix = "arXiv",
    primaryClass = "math.AG",
    doi = "10.4310/MRL.1998.v5.n4.a7",
    journal = "Math. Res. Lett.",
    volume = "5",
    pages = "497--516",
    year = "1998"
}

@article{vonManteuffel:2014qoa,
    author = "von Manteuffel, Andreas and Panzer, Erik and Schabinger, Robert M.",
    title = "{A quasi-finite basis for multi-loop Feynman integrals}",
    eprint = "1411.7392",
    archivePrefix = "arXiv",
    primaryClass = "hep-ph",
    reportNumber = "MITP-14-076",
    doi = "10.1007/JHEP02(2015)120",
    journal = "JHEP",
    volume = "02",
    pages = "120",
    year = "2015"
}

@article{Beneke:2002ph,
    author = "Beneke, M. and Chapovsky, A. P. and Diehl, M. and Feldmann, T.",
    title = "{Soft collinear effective theory and heavy to light currents beyond leading power}",
    eprint = "hep-ph/0206152",
    archivePrefix = "arXiv",
    reportNumber = "PITHA-02-09",
    doi = "10.1016/S0550-3213(02)00687-9",
    journal = "Nucl. Phys. B",
    volume = "643",
    pages = "431--476",
    year = "2002"
}

@article{Chetyrkin:2003vi,
    author = "Chetyrkin, K. G. and Grozin, A. G.",
    title = "{Three loop anomalous dimension of the heavy light quark current in HQET}",
    eprint = "hep-ph/0303113",
    archivePrefix = "arXiv",
    reportNumber = "TTP03-10, SFB-CPP-03-05, THEP-03-05",
    doi = "10.1016/S0550-3213(03)00490-5",
    journal = "Nucl. Phys. B",
    volume = "666",
    pages = "289--302",
    year = "2003"
}

@article{Wang:2016qii,
    author = "Wang, Yu-Ming",
    title = "{Factorization and dispersion relations for radiative leptonic $B$ decay}",
    eprint = "1606.03080",
    archivePrefix = "arXiv",
    primaryClass = "hep-ph",
    reportNumber = "UWTHPH-2016-10",
    doi = "10.1007/JHEP09(2016)159",
    journal = "JHEP",
    volume = "09",
    pages = "159",
    year = "2016"
}

@article{Becher:2009kw,
    author = "Becher, Thomas and Neubert, Matthias",
    title = "{Infrared singularities of QCD amplitudes with massive partons}",
    eprint = "0904.1021",
    archivePrefix = "arXiv",
    primaryClass = "hep-ph",
    reportNumber = "FERMILAB-PUB-09-099-T, MZ-TH-09-11",
    doi = "10.1103/PhysRevD.79.125004",
    journal = "Phys. Rev. D",
    volume = "79",
    pages = "125004",
    year = "2009",
    note = "[Erratum: Phys.Rev.D 80, 109901 (2009)]"
}

@article{Gehrmann:1999as,
    author = "Gehrmann, T. and Remiddi, E.",
    title = "{Differential equations for two-loop four-point functions}",
    eprint = "hep-ph/9912329",
    archivePrefix = "arXiv",
    reportNumber = "TTP-99-49",
    doi = "10.1016/S0550-3213(00)00223-6",
    journal = "Nucl. Phys. B",
    volume = "580",
    pages = "485--518",
    year = "2000"
}

@article{Meyer:2017joq,
    author = "Meyer, Christoph",
    title = "{Algorithmic transformation of multi-loop master integrals to a canonical basis with CANONICA}",
    eprint = "1705.06252",
    archivePrefix = "arXiv",
    primaryClass = "hep-ph",
    reportNumber = "HU-EP-17-10",
    doi = "10.1016/j.cpc.2017.09.014",
    journal = "Comput. Phys. Commun.",
    volume = "222",
    pages = "295--312",
    year = "2018"
}

@article{Lunghi:2002ju,
    author = "Lunghi, Enrico and Pirjol, Dan and Wyler, Daniel",
    title = "{Factorization in leptonic radiative B $\to \gamma$e{\ensuremath{\nu}} decays}",
    eprint = "hep-ph/0210091",
    archivePrefix = "arXiv",
    reportNumber = "ZU-TH-17-02",
    doi = "10.1016/S0550-3213(02)01032-5",
    journal = "Nucl. Phys. B",
    volume = "649",
    pages = "349--364",
    year = "2003"
}

@article{Bauer:2000ew,
    author = "Bauer, Christian W. and Fleming, Sean and Luke, Michael E.",
    title = "{Summing Sudakov logarithms in $B \to  X_s \gamma $in effective field theory.}",
    eprint = "hep-ph/0005275",
    archivePrefix = "arXiv",
    reportNumber = "UTPT-00-03",
    doi = "10.1103/PhysRevD.63.014006",
    journal = "Phys. Rev. D",
    volume = "63",
    pages = "014006",
    year = "2000"
}

@article{Bauer:2000yr,
    author = "Bauer, Christian W. and Fleming, Sean and Pirjol, Dan and Stewart, Iain W.",
    title = "{An Effective field theory for collinear and soft gluons: Heavy to light decays}",
    eprint = "hep-ph/0011336",
    archivePrefix = "arXiv",
    reportNumber = "UCSD-PTH-00-28",
    doi = "10.1103/PhysRevD.63.114020",
    journal = "Phys. Rev. D",
    volume = "63",
    pages = "114020",
    year = "2001"
}

@article{Braun:2019wyx,
    author = "Braun, V. M. and Ji, Yao and Manashov, A. N.",
    title = "{Two-loop evolution equation for the B-meson distribution amplitude}",
    eprint = "1905.04498",
    archivePrefix = "arXiv",
    primaryClass = "hep-ph",
    reportNumber = "DESY-19-080",
    doi = "10.3204/PUBDB-2019-02451",
    journal = "Phys. Rev. D",
    volume = "100",
    number = "1",
    pages = "014023",
    year = "2019"
}

@article{Lange:2003ff,
    author = "Lange, Bjorn O. and Neubert, Matthias",
    title = "{Renormalization group evolution of the B meson light cone distribution amplitude}",
    eprint = "hep-ph/0303082",
    archivePrefix = "arXiv",
    reportNumber = "CLNS-03-1822",
    doi = "10.1103/PhysRevLett.91.102001",
    journal = "Phys. Rev. Lett.",
    volume = "91",
    pages = "102001",
    year = "2003"
}

@article{Moch:2004pa,
    author = "Moch, S. and Vermaseren, J. A. M. and Vogt, A.",
    title = "{The Three loop splitting functions in QCD: The Nonsinglet case}",
    eprint = "hep-ph/0403192",
    archivePrefix = "arXiv",
    reportNumber = "DESY-04-047, SFB-CPP-04-09, NIKHEF-04-001",
    doi = "10.1016/j.nuclphysb.2004.03.030",
    journal = "Nucl. Phys. B",
    volume = "688",
    pages = "101--134",
    year = "2004"
}

@article{Beneke:2000ry,
    author = "Beneke, M. and Buchalla, G. and Neubert, M. and Sachrajda, Christopher T.",
    title = "{QCD factorization for exclusive, nonleptonic B meson decays: General arguments and the case of heavy light final states}",
    eprint = "hep-ph/0006124",
    archivePrefix = "arXiv",
    reportNumber = "CERN-TH-2000-159, CLNS-00-1675, PITHA-00-06, SHEP-00-06",
    doi = "10.1016/S0550-3213(00)00559-9",
    journal = "Nucl. Phys. B",
    volume = "591",
    pages = "313--418",
    year = "2000"
}

@article{Grozin:2015kna,
    author = "Grozin, Andrey and Henn, Johannes M. and Korchemsky, Gregory P. and Marquard, Peter",
    title = "{The three-loop cusp anomalous dimension in QCD and its supersymmetric extensions}",
    eprint = "1510.07803",
    archivePrefix = "arXiv",
    primaryClass = "hep-ph",
    reportNumber = "DESY-15-186, IPHT-T15-174, MITP-15-088",
    doi = "10.1007/JHEP01(2016)140",
    journal = "JHEP",
    volume = "01",
    pages = "140",
    year = "2016"
}

@article{Arkani-Hamed:2010pyv,
    author = "Arkani-Hamed, Nima and Bourjaily, Jacob L. and Cachazo, Freddy and Trnka, Jaroslav",
    title = "{Local Integrals for Planar Scattering Amplitudes}",
    eprint = "1012.6032",
    archivePrefix = "arXiv",
    primaryClass = "hep-th",
    doi = "10.1007/JHEP06(2012)125",
    journal = "JHEP",
    volume = "06",
    pages = "125",
    year = "2012"
}

@article{Grozin:2014hna,
    author = "Grozin, Andrey and Henn, Johannes M. and Korchemsky, Gregory P. and Marquard, Peter",
    title = "{Three Loop Cusp Anomalous Dimension in QCD}",
    eprint = "1409.0023",
    archivePrefix = "arXiv",
    primaryClass = "hep-ph",
    reportNumber = "IPHT-T14-111, DESY-14-148, SFB-CPP-14-64, LPN14-104",
    doi = "10.1103/PhysRevLett.114.062006",
    journal = "Phys. Rev. Lett.",
    volume = "114",
    number = "6",
    pages = "062006",
    year = "2015"
}

@article{Bruser:2019yjk,
    author = {Br{\"u}ser, Robin and Liu, Ze Long and Stahlhofen, Maximilian},
    title = "{Three-loop soft function for heavy-to-light quark decays}",
    eprint = "1911.04494",
    archivePrefix = "arXiv",
    primaryClass = "hep-ph",
    reportNumber = "MITP/19-073, P3H-19-042, SI-HEP-2019-15",
    doi = "10.1007/JHEP03(2020)071",
    journal = "JHEP",
    volume = "03",
    pages = "071",
    year = "2020"
}

@article{Becher:2009qa,
    author = "Becher, Thomas and Neubert, Matthias",
    title = "{On the Structure of Infrared Singularities of Gauge-Theory Amplitudes}",
    eprint = "0903.1126",
    archivePrefix = "arXiv",
    primaryClass = "hep-ph",
    reportNumber = "FERMILAB-PUB-09-061-T, MZ-TH-09-05",
    doi = "10.1088/1126-6708/2009/06/081",
    journal = "JHEP",
    volume = "06",
    pages = "081",
    year = "2009",
    note = "[Erratum: JHEP 11, 024 (2013)]"
}

@article{Beneke:2011nf,
    author = "Beneke, M. and Rohrwild, J.",
    title = "{B meson distribution amplitude from B --{\ensuremath{>}} {\textbackslash}gamma l {\textbackslash}nu}",
    eprint = "1110.3228",
    archivePrefix = "arXiv",
    primaryClass = "hep-ph",
    reportNumber = "TTK-11-51, SFB-CPP-11-56",
    doi = "10.1140/epjc/s10052-011-1818-8",
    journal = "Eur. Phys. J. C",
    volume = "71",
    pages = "1818",
    year = "2011"
}

@article{Braun:2014owa,
    author = "Braun, V. M. and Manashov, A. N.",
    title = "{Conformal symmetry of the Lange-Neubert evolution equation}",
    eprint = "1402.5822",
    archivePrefix = "arXiv",
    primaryClass = "hep-ph",
    doi = "10.1016/j.physletb.2014.02.051",
    journal = "Phys. Lett. B",
    volume = "731",
    pages = "316--319",
    year = "2014"
}

@article{Wang:2019msf,
    author = "Wang, Wei and Wang, Yu-Ming and Xu, Ji and Zhao, Shuai",
    title = "{$B$-meson light-cone distribution amplitude from Euclidean quantities}",
    eprint = "1908.09933",
    archivePrefix = "arXiv",
    primaryClass = "hep-ph",
    reportNumber = "JLAB-THY-19-3025",
    doi = "10.1103/PhysRevD.102.011502",
    journal = "Phys. Rev. D",
    volume = "102",
    number = "1",
    pages = "011502",
    year = "2020"
}

@article{Braun:2003wx,
    author = "Braun, V. M. and Ivanov, D. Yu. and Korchemsky, G. P.",
    title = "{The B meson distribution amplitude in QCD}",
    eprint = "hep-ph/0309330",
    archivePrefix = "arXiv",
    reportNumber = "LPT-ORSAY-03-63",
    doi = "10.1103/PhysRevD.69.034014",
    journal = "Phys. Rev. D",
    volume = "69",
    pages = "034014",
    year = "2004"
}

@article{Moch:2005id,
    author = "Moch, S. and Vermaseren, J. A. M. and Vogt, A.",
    title = "{The Quark form-factor at higher orders}",
    eprint = "hep-ph/0507039",
    archivePrefix = "arXiv",
    reportNumber = "SFB-CPP-05-26, DCPT-05-66, IPP-05-33, NIKHEF-05-011, DESY-05-106",
    doi = "10.1088/1126-6708/2005/08/049",
    journal = "JHEP",
    volume = "08",
    pages = "049",
    year = "2005"
}

@article{Henn:2013pwa,
    author = "Henn, Johannes M.",
    title = "{Multiloop integrals in dimensional regularization made simple}",
    eprint = "1304.1806",
    archivePrefix = "arXiv",
    primaryClass = "hep-th",
    doi = "10.1103/PhysRevLett.110.251601",
    journal = "Phys. Rev. Lett.",
    volume = "110",
    pages = "251601",
    year = "2013"
}

@article{Beneke:2018wjp,
    author = "Beneke, M. and Braun, V. M. and Ji, Yao and Wei, Yan-Bing",
    title = "{Radiative leptonic decay $B\to \gamma \ell \nu_\ell$ with subleading power corrections}",
    eprint = "1804.04962",
    archivePrefix = "arXiv",
    primaryClass = "hep-ph",
    reportNumber = "TUM-HEP-1135-18",
    doi = "10.1007/JHEP07(2018)154",
    journal = "JHEP",
    volume = "07",
    pages = "154",
    year = "2018"
}

@article{Panzer:2014caa,
    author = "Panzer, Erik",
    title = "{Algorithms for the symbolic integration of hyperlogarithms with applications to Feynman integrals}",
    eprint = "1403.3385",
    archivePrefix = "arXiv",
    primaryClass = "hep-th",
    doi = "10.1016/j.cpc.2014.10.019",
    journal = "Comput. Phys. Commun.",
    volume = "188",
    pages = "148--166",
    year = "2015"
}

@article{Baikov:1996rk,
    author = "Baikov, P. A.",
    title = "{Explicit solutions of the three loop vacuum integral recurrence relations}",
    eprint = "hep-ph/9603267",
    archivePrefix = "arXiv",
    reportNumber = "INP-96-10-417",
    doi = "10.1016/0370-2693(96)00835-0",
    journal = "Phys. Lett. B",
    volume = "385",
    pages = "404--410",
    year = "1996"
}

@article{Bauer:2001ct,
    author = "Bauer, Christian W. and Stewart, Iain W.",
    title = "{Invariant operators in collinear effective theory}",
    eprint = "hep-ph/0107001",
    archivePrefix = "arXiv",
    reportNumber = "UCSD-PTH-01-09",
    doi = "10.1016/S0370-2693(01)00902-9",
    journal = "Phys. Lett. B",
    volume = "516",
    pages = "134--142",
    year = "2001"
}

@article{Bauer:2001yt,
    author = "Bauer, Christian W. and Pirjol, Dan and Stewart, Iain W.",
    title = "{Soft collinear factorization in effective field theory}",
    eprint = "hep-ph/0109045",
    archivePrefix = "arXiv",
    reportNumber = "UCSD-PTH-01-15",
    doi = "10.1103/PhysRevD.65.054022",
    journal = "Phys. Rev. D",
    volume = "65",
    pages = "054022",
    year = "2002"
}

@article{Liu:2020wbn,
    author = "Liu, Ze Long and Mecaj, Bianka and Neubert, Matthias and Wang, Xing",
    title = "{Factorization at subleading power and endpoint divergences in $h\to\gamma\gamma$ decay. Part II. Renormalization and scale evolution}",
    eprint = "2009.06779",
    archivePrefix = "arXiv",
    primaryClass = "hep-ph",
    reportNumber = "MITP/20-047",
    doi = "10.1007/JHEP01(2021)077",
    journal = "JHEP",
    volume = "01",
    pages = "077",
    year = "2021"
}

@article{Liu:2020tzd,
    author = "Liu, Ze Long and Mecaj, Bianka and Neubert, Matthias and Wang, Xing",
    title = "{Factorization at subleading power, Sudakov resummation, and endpoint divergences in soft-collinear effective theory}",
    eprint = "2009.04456",
    archivePrefix = "arXiv",
    primaryClass = "hep-ph",
    reportNumber = "MITP/20-050, ZU-TH 31/20",
    doi = "10.1103/PhysRevD.104.014004",
    journal = "Phys. Rev. D",
    volume = "104",
    number = "1",
    pages = "014004",
    year = "2021"
}

@article{LPC:2026vyv,
    author = "Gao, Hao-Fei and others",
    collaboration = "LPC",
    title = "{Determination of heavy meson light-cone distribution amplitudes: theoretical framework and lattice simulations}",
    eprint = "2604.25802",
    archivePrefix = "arXiv",
    primaryClass = "hep-lat",
    month = "4",
    year = "2026"
}

@article{Ferre:2026sao,
    author = "Ferr{\'e}, Max and Neubert, Matthias",
    title = "{Electromagnetic tails of $B$-meson light-cone distribution amplitudes}",
    eprint = "2609.01718",
    archivePrefix = "arXiv",
    primaryClass = "hep-ph",
    month = "9",
    year = "2026"
}

@article{Asatrian:2008uk,
    author = "Asatrian, H. M. and Greub, C. and Pecjak, B. D.",
    title = "{NNLO corrections to anti-B ---{\ensuremath{>}} X(u) l anti-nu in the shape-function region}",
    eprint = "0810.0987",
    archivePrefix = "arXiv",
    primaryClass = "hep-ph",
    reportNumber = "DESY-08-145",
    doi = "10.1103/PhysRevD.78.114028",
    journal = "Phys. Rev. D",
    volume = "78",
    pages = "114028",
    year = "2008"
}

@article{Beneke:2008ei,
    author = "Beneke, M. and Huber, T. and Li, X. -Q.",
    title = "{Two-loop QCD correction to differential semi-leptonic b ---{\ensuremath{>}} u decays in the shape-function region}",
    eprint = "0810.1230",
    archivePrefix = "arXiv",
    primaryClass = "hep-ph",
    reportNumber = "PITHA-08-25, SFB-CPP-08-80",
    doi = "10.1016/j.nuclphysb.2008.11.019",
    journal = "Nucl. Phys. B",
    volume = "811",
    pages = "77--97",
    year = "2009"
}

@article{Bell:2008ws,
    author = "Bell, Guido",
    title = "{NNLO corrections to inclusive semileptonic B decays in the shape-function region}",
    eprint = "0810.5695",
    archivePrefix = "arXiv",
    primaryClass = "hep-ph",
    reportNumber = "TTP08-49, SFB-CPP-08-90",
    doi = "10.1016/j.nuclphysb.2008.12.018",
    journal = "Nucl. Phys. B",
    volume = "812",
    pages = "264--289",
    year = "2009"
}

@article{Grozin:2005iz,
    author = "Grozin, Andrey G.",
    title = "{B-meson distribution amplitudes}",
    eprint = "hep-ph/0506226",
    archivePrefix = "arXiv",
    doi = "10.1142/S0217751X05028570",
    journal = "Int. J. Mod. Phys. A",
    volume = "20",
    pages = "7451--7484",
    year = "2005"
}

@article{Laenen:2020nrt,
    author = "Laenen, E. and Sinninghe Damst{\'e}, J. and Vernazza, L. and Waalewijn, W. and Zoppi, L.",
    title = "{Towards all-order factorization of QED amplitudes at next-to-leading power}",
    eprint = "2008.01736",
    archivePrefix = "arXiv",
    primaryClass = "hep-ph",
    reportNumber = "NIKHEF 2020-019",
    doi = "10.1103/PhysRevD.103.034022",
    journal = "Phys. Rev. D",
    volume = "103",
    number = "3",
    pages = "034022",
    year = "2021"
}

@article{Beneke:2022obx,
    author = "Beneke, M. and Garny, M. and Jaskiewicz, S. and Strohm, J. and Szafron, R. and Vernazza, L. and Wang, J.",
    title = "{Next-to-leading power endpoint factorization and resummation for off-diagonal {\textquotedblleft}gluon{\textquotedblright} thrust}",
    eprint = "2205.04479",
    archivePrefix = "arXiv",
    primaryClass = "hep-ph",
    reportNumber = "IPPP/22/25, TUM-HEP-1398/22",
    doi = "10.1007/JHEP07(2022)144",
    journal = "JHEP",
    volume = "07",
    pages = "144",
    year = "2022"
}

@article{Beneke:2017ztn,
    author = "Beneke, Martin and Garny, Mathias and Szafron, Robert and Wang, Jian",
    title = "{Anomalous dimension of subleading-power N-jet operators}",
    eprint = "1712.04416",
    archivePrefix = "arXiv",
    primaryClass = "hep-ph",
    reportNumber = "TUM-HEP-1116-17",
    doi = "10.1007/JHEP03(2018)001",
    journal = "JHEP",
    volume = "03",
    pages = "001",
    year = "2018"
}

@article{Beneke:2024cpq,
    author = "Beneke, Martin and Ji, Yao and Wang, Xing",
    title = "{Renormalization of the next-to-leading-power {\ensuremath{\gamma}}{\ensuremath{\gamma}} {\textrightarrow} h and gg {\textrightarrow} h soft quark functions}",
    eprint = "2403.17738",
    archivePrefix = "arXiv",
    primaryClass = "hep-ph",
    reportNumber = "TUM-HEP-1501/24",
    doi = "10.1007/JHEP05(2024)246",
    journal = "JHEP",
    volume = "05",
    pages = "246",
    year = "2024"
}

@article{Moult:2019mog,
    author = "Moult, Ian and Stewart, Iain W. and Vita, Gherardo",
    title = "{Subleading Power Factorization with Radiative Functions}",
    eprint = "1905.07411",
    archivePrefix = "arXiv",
    primaryClass = "hep-ph",
    reportNumber = "MIT-CTP 4948",
    doi = "10.1007/JHEP11(2019)153",
    journal = "JHEP",
    volume = "11",
    pages = "153",
    year = "2019"
}

@article{Braun:2017liq,
    author = "Braun, V. M. and Ji, Yao and Manashov, A. N.",
    title = "{Higher-twist B-meson Distribution Amplitudes in HQET}",
    eprint = "1703.02446",
    archivePrefix = "arXiv",
    primaryClass = "hep-ph",
    reportNumber = "DESY-17-037",
    doi = "10.1007/JHEP05(2017)022",
    journal = "JHEP",
    volume = "05",
    pages = "022",
    year = "2017"
}

@article{Korchemsky:1999qb,
    author = "Korchemsky, Gregory P. and Pirjol, Dan and Yan, Tung-Mow",
    title = "{Radiative leptonic decays of B mesons in QCD}",
    eprint = "hep-ph/9911427",
    archivePrefix = "arXiv",
    reportNumber = "CLNS-99-1644, LPT-ORSAY-99-93, UCSD-PTH-99-17",
    doi = "10.1103/PhysRevD.61.114510",
    journal = "Phys. Rev. D",
    volume = "61",
    pages = "114510",
    year = "2000"
}

@article{Bell:2010mg,
    author = "Bell, G. and Beneke, M. and Huber, T. and Li, Xin-Qiang",
    title = "{Heavy-to-light currents at NNLO in SCET and semi-inclusive $\bar{B} \to  X_s l^+ l^-$ decay}",
    eprint = "1007.3758",
    archivePrefix = "arXiv",
    primaryClass = "hep-ph",
    reportNumber = "TTK-10-42, SI-HEP-2010-10, TTP10-32, SFB-CPP-10-63",
    doi = "10.1016/j.nuclphysb.2010.09.022",
    journal = "Nucl. Phys. B",
    volume = "843",
    pages = "143--176",
    year = "2011"
}

@article{Huang:2026fck,
    author = "Huang, Yong-Kang and Ji, Yao and Shi, Bo-Xuan and Wang, Xing and Wang, Yu-Ming and Wei, Yu-Xiang",
    title = "{Next-to-Next-to-Next-to-Leading-Order Prediction for the Radiative $B_{u}^- \to γ\ell^{-} \bar ν_{\ell}$ Decay}",
    eprint = "2611.18202",
    archivePrefix = "arXiv",
    primaryClass = "hep-ph",
    month = "11",
    year = "2026"
}

@article{Ji:2026jwu,
    author = "Ji, Yao and Shi, Bo-Xuan and Wang, Xing and Wang, Yu-Ming",
    title = "{Three-Loop Evolution for Soft Functions of Exclusive $B$-Meson and Higgs Decays}",
    eprint = "2611.18204",
    archivePrefix = "arXiv",
    primaryClass = "hep-ph",
    month = "11",
    year = "2026"
}

@article{LHCb:2026clt,
    author = "Aaij, Roel and others",
    collaboration = "LHCb",
    title = "{Search for the decays $B^+_{(c)} \to μ^+ ν_μγ$}",
    eprint = "2607.12825",
    archivePrefix = "arXiv",
    primaryClass = "hep-ex",
    reportNumber = "LHCb-PAPER-2026-020,CERN-EP-2026-179",
    month = "7",
    year = "2026"
}

@article{Fael:2024vko,
    author = {Fael, Matteo and Huber, Tobias and Lange, Fabian and M{\"u}ller, Jakob and Sch{\"o}nwald, Kay and Steinhauser, Matthias},
    title = "{Heavy-to-light form factors to three loops}",
    eprint = "2406.08182",
    archivePrefix = "arXiv",
    primaryClass = "hep-ph",
    reportNumber = "CERN-TH-2024-064, P3H-24-036, PSI-PR-24-13, SI-HEP-2024-14, TTP24-017, SI-HEP-2024-14,
  TTP24-017, ZU-TH28/24",
    doi = "10.1103/PhysRevD.110.056011",
    journal = "Phys. Rev. D",
    volume = "110",
    number = "5",
    pages = "056011",
    year = "2024"
}

@article{Bekavac:2009zc,
    author = "Bekavac, S. and Grozin, A. G. and Marquard, P. and Piclum, J. H. and Seidel, D. and Steinhauser, M.",
    title = "{Matching QCD and HQET heavy-light currents at three loops}",
    eprint = "0911.3356",
    archivePrefix = "arXiv",
    primaryClass = "hep-ph",
    reportNumber = "TTP09-41, SFB-CPP-09-110, ALBERTA-THY-16-09",
    doi = "10.1016/j.nuclphysb.2010.02.025",
    journal = "Nucl. Phys. B",
    volume = "833",
    pages = "46--63",
    year = "2010"
}

@article{Bazavov:2017lyh,
    author = "Bazavov, A. and others",
    title = "{$B$- and $D$-meson leptonic decay constants from four-flavor lattice QCD}",
    eprint = "1712.09262",
    archivePrefix = "arXiv",
    primaryClass = "hep-lat",
    reportNumber = "FERMILAB-PUB-17/491-T, FERMILAB-PUB-17-491-T",
    doi = "10.1103/PhysRevD.98.074512",
    journal = "Phys. Rev. D",
    volume = "98",
    number = "7",
    pages = "074512",
    year = "2018"
}

@article{Belle-II:2024xwh,
    author = "Adachi, I. and others",
    collaboration = "Belle-II",
    title = "{Determination of |Vub| from simultaneous measurements of untagged B0{\textrightarrow}{\ensuremath{\pi}}-{\ensuremath{\ell}}+{\ensuremath{\nu}}{\ensuremath{\ell}} and B+{\textrightarrow}{\ensuremath{\rho}}0{\ensuremath{\ell}}+{\ensuremath{\nu}}{\ensuremath{\ell}} decays}",
    eprint = "2407.17403",
    archivePrefix = "arXiv",
    primaryClass = "hep-ex",
    reportNumber = "Belle II Preprint 2024-023, KEK Preprint 2024-21",
    doi = "10.1103/jnwn-ts6q",
    journal = "Phys. Rev. D",
    volume = "111",
    number = "11",
    pages = "112009",
    year = "2025"
}

@article{ParticleDataGroup:2024cfk,
    author = "Navas, S. and others",
    collaboration = "Particle Data Group",
    title = "{Review of particle physics}",
    doi = "10.1103/PhysRevD.110.030001",
    journal = "Phys. Rev. D",
    volume = "110",
    number = "3",
    pages = "030001",
    year = "2024"
}

@article{Liu:2019oav,
    author = "Liu, Ze Long and Neubert, Matthias",
    title = "{Factorization at subleading power and endpoint-divergent convolutions in $h\to\gamma\gamma$ decay}",
    eprint = "1912.08818",
    archivePrefix = "arXiv",
    primaryClass = "hep-ph",
    reportNumber = "MITP/19-078",
    doi = "10.1007/JHEP04(2020)033",
    journal = "JHEP",
    volume = "04",
    pages = "033",
    year = "2020"
}

@article{Bonciani:2008wf,
    author = "Bonciani, R. and Ferroglia, A.",
    title = "{Two-Loop QCD Corrections to the Heavy-to-Light Quark Decay}",
    eprint = "0809.4687",
    archivePrefix = "arXiv",
    primaryClass = "hep-ph",
    reportNumber = "ZU-TH-16-08",
    doi = "10.1088/1126-6708/2008/11/065",
    journal = "JHEP",
    volume = "11",
    pages = "065",
    year = "2008"
}

@article{Braun:2012kp,
    author = "Braun, V. M. and Khodjamirian, A.",
    title = "{Soft contribution to $B\to \gamma \ell \nu_\ell$ and the $B$-meson distribution amplitude}",
    eprint = "1210.4453",
    archivePrefix = "arXiv",
    primaryClass = "hep-ph",
    reportNumber = "SI-HEP-2012-19",
    doi = "10.1016/j.physletb.2012.11.047",
    journal = "Phys. Lett. B",
    volume = "718",
    pages = "1014--1019",
    year = "2013"
}

@article{Gehrmann:2010ue,
    author = "Gehrmann, T. and Glover, E. W. N. and Huber, T. and Ikizlerli, N. and Studerus, C.",
    title = "{Calculation of the quark and gluon form factors to three loops in QCD}",
    eprint = "1004.3653",
    archivePrefix = "arXiv",
    primaryClass = "hep-ph",
    reportNumber = "ZU-TH-07-10, IPPP-10-26, SI-HEP-2010-09, ZU--TH-07-10, ZU--TH 07/10, IPPP/10/26, SI-HEP-2010-09",
    doi = "10.1007/JHEP06(2010)094",
    journal = "JHEP",
    volume = "06",
    pages = "094",
    year = "2010"
}


\onecolumngrid
\newpage
\appendix

\section*{Supplemental Material}
\allowdisplaybreaks[2]

\subsection{A. Three-loop Lange-Neubert kernel}\label{app:kernel}
In this section, we present the explicit expressions of the three-loop Lange-Neubert kernel $\gamma_+(\omega,\omega',\mu)$ in~\eqref{eq:gamma3}. The perturbative expansions of the anomalous dimensions are defined as
\begin{gather}
\Gamma_{\rm cusp}(\alpha_s)=\sum_{n\ge0}\Gamma_n\Big(\frac{\alpha_s}{4\pi}\Big)^{n+1}\,,\qquad
\gamma(\alpha_s) = \sum_{n\ge0}\gamma_n\Big(\frac{\alpha_s}{4\pi}\Big)^{n+1}\,,
\end{gather}
and similarly for all other anomalous dimensions, e.g. for $\Gamma_{\rm hl}$, $\gamma^q$, $\gamma^Q$, $\gamma_F$ and $\gamma'$, in the following discussion.

The one-, two-, and three-loop coefficients of the cusp anomalous dimension are~\cite{Korchemsky:1987wg,Moch:2004pa}
\begin{align} 
\Gamma_0 ={}& 4C_F
\,,
\label{eq:Gcusp0}\\
\Gamma_1 ={}& 4C_F \bigg[\biggl( \frac{67}{9} -\frac{\pi^2}{3} \biggr)\,C_A  -
   \frac{20}{9}\,n_f T_F \bigg]
\,,
\label{eq:Gcusp1}\\
\Gamma_2 ={}& 4C_F \bigg[
\biggl(\frac{245}{6} -\frac{134 \pi^2}{27} + \frac{11 \pi ^4}{45}
  + \frac{22 \zeta_3}{3}\biggr)C_A^2
  + \biggl(- \frac{418}{27} + \frac{40 \pi^2}{27}  - \frac{56 \zeta_3}{3} 
\biggr)C_A n_f T_F
\nn\\
&\quad  + \biggl(- \frac{55}{3} + 16 \zeta_3 \biggr) C_F n_f T_F
  - \frac{16}{27}\, (n_f T_F)^2 \bigg]\,.
\label{eq:Gcusp2}
\end{align}
The coefficients of the non-cusp anomalous dimension $\gamma(\alpha_s)$ in~\eqref{eq:gamma3} are
\begin{align}
\gamma_0={}&-2C_F\,,\label{eq:gamma0}\\
\gamma_1={}&C_F^2\left(-4+\frac{14\pi^2}{3}-24\zeta_3\right)
+C_FC_A\left(\frac{254}{27}-\frac{55\pi^2}{18}-6\zeta_3\right)
+C_FT_F n_f\left(-\frac{64}{27}+\frac{10\pi^2}{9}\right),\label{eq:gamma1}\\
\gamma_2={}&C_F^3\left(4-\frac{41\pi^2}{3}-\frac{32\pi^4}{45}-32\zeta_3+\frac{16\pi^2}{3}\zeta_3+240\zeta_5\right)\nn\\
&+C_F^2C_A\left(-\frac{176}{9}+\frac{1207\pi^2}{27}+\frac{271\pi^4}{135}-\frac{986}{3}\zeta_3-\frac{8\pi^2}{3}\zeta_3-120\zeta_5\right)\nn\\
&+C_FC_A^2\left(\frac{128735}{1458}-\frac{16681\pi^2}{486}+\frac{73\pi^4}{90}-\frac{1096}{9}\zeta_3+\frac{32\pi^2}{9}\zeta_3+92\zeta_5\right)\nn\\
&+C_F^2T_F n_f\left(-\frac{889}{27}-\frac{266\pi^2}{27}-\frac{92\pi^4}{135}+\frac{1360}{9}\zeta_3\right)\nn\\
&+C_FC_AT_F n_f\left(-\frac{17566}{729}+\frac{5206\pi^2}{243}-\frac{26\pi^4}{45}-\frac{352}{27}\zeta_3\right)
+C_FT_F^2n_f^2\left(-\frac{1568}{729}-\frac{200\pi^2}{81}+\frac{416}{27}\zeta_3\right).\label{eq:gamma2}
\end{align}
The functions $h^{(n)}(x)$ in~\eqref{eq:gamma3} are real for $x=\omega/\omega'<1$ and are given by
\begin{align}
 h^{(2)}(x)={}&-4C_F\ln x\left[\beta_0+2C_F\left(\ln x-\frac32-\frac{1+x}{x}\ln(1-x)\right)\right]
\label{eq:h2classic}
  \\
 h^{(3)}(x)={}&C_F^3\ln x\Bigg\{-4\ln^3x + \left[18+12\frac{1+x}{x}\ln(1-x)\right]\ln^2 x 
 +24\frac{1+x}{x}\left[\zeta_3-\mathrm{Li}_3(1-x)+\mathrm{Li}_3(x)\right]
 \nn\\
 &\qquad
 +\left[-50+\frac{8\pi^2}{3}-24\frac{1+x}{x}\ln(1-x) -20\frac{1+x}{x}\ln^2(1-x)-16\mathrm{Li}_2(x)\right]\ln x
 \nn\\
 &\qquad+6-4\pi^2-\frac{24}{x}\mathrm{Li}_2(x) +\left[\frac{4(1-x)}x+\frac{4\pi^2(1+x)}{3x}
-8\frac{1+x}{x}\mathrm{Li}_2(x)\right]\ln(1-x)
\Bigg\}
 \nn\\
 &+C_F^2C_A\ln x\Bigg\{-22\ln^2x
 +\left[-\frac{124}{3}+4\pi^2+\frac{88(1+x)}{3x}\ln(1-x)
 +8\frac{1+x}{x}\ln^2(1-x)\right]\ln x
 \nn\\
 &\qquad+\frac{302}{3}+\frac{8\pi^2}{9}+\frac{88}{3x}\mathrm{Li}_2(x) +\left[\frac{4(259+277x)}{9x}-\frac{20\pi^2(1+x)}{3x}
 +8\frac{1+x}{x}\mathrm{Li}_2(x)\right]\ln(1-x)
 \nn\\
 &\qquad+24\frac{1+x}{x}\left[\mathrm{Li}_3(1-x)-\zeta_3\right]\Bigg\}
 \nn\\
 &+C_FC_A^2\ln x\left[-\frac{242}{9}\ln x+\frac{4}{27}(-1043+33\pi^2)\right]
 \nn\\
 &+C_F^2 T_F n_f\ln x\Bigg\{8\ln^2x + \left[\frac{32}{3}-\frac{32(1+x)}{3x}\ln(1-x)\right]\ln x
-\frac{40}{3}-\frac{16\pi^2}{9}-\frac{320(1+x)}{9x}\ln(1-x)
 -\frac{32}{3x}\mathrm{Li}_2(x)\Bigg\} 
  \nn\\
 & 
 + C_F C_A T_F n_f\ln x\left[\frac{176}{9}\ln x+\frac{16}{27}(167-3\pi^2)\right]
 +C_F T_F^2 n_f^2\ln x\left(-\frac{32}{9}\ln x-\frac{320}{27}\right)\,.
 \label{eq:h3classic}
\end{align}
The two-loop function $h^{(2)}(x)$ agrees with~\cite{Liu:2020ydl}. Although harmonic polylogarithms of weight four appear in intermediate expressions, $\mathrm{Li}_4$ and the non-classical polylogarithms cancel in $h^{(3)}(x)$.

\subsection{B. Infrared renormalization factor}\label{app:IR}
The IR renormalization factor $Z_{\rm hl}$, which removes the IR poles of the partonic matrix elements in~\eqref{eq:finite}, is determined by the anomalous dimension $\Gamma_{\rm hl}$. Solving~\eqref{eq:ZhlRGE} for $Z_{\rm hl}$ order by order in $\alpha_s$~\cite{Becher:2009qa}, we obtain
\begin{align}\label{eq:Zhl}
\ln Z_{\rm hl}(\epsilon,\omega',\mu)&=\frac{\alpha_s}{4\pi}\left(-\frac{\Gamma_0}{4\epsilon^2}+\frac{\Gamma_0^{\rm hl}}{2\epsilon}\right)
+\Big(\frac{\alpha_s}{4\pi}\Big)^2\left(\frac{3\beta_0\Gamma_0}{16\epsilon^3}-\frac{\Gamma_1+4\beta_0\Gamma_0^{\rm hl}}{16\epsilon^2}+\frac{\Gamma_1^{\rm hl}}{4\epsilon}\right) \nn\\
&\quad+\Big(\frac{\alpha_s}{4\pi}\Big)^3\left(-\frac{11\beta_0^2\Gamma_0}{72\epsilon^4}+\frac{5\beta_0\Gamma_1+8\beta_1\Gamma_0+12\beta_0^2\Gamma_0^{\rm hl}}{72\epsilon^3}-\frac{\Gamma_2+6\beta_0\Gamma_1^{\rm hl}+6\beta_1\Gamma_0^{\rm hl}}{36\epsilon^2}+\frac{\Gamma_2^{\rm hl}}{6\epsilon}\right)+\mathcal{O}(\alpha_s^4)\,,
\end{align}
with
\begin{equation}
\beta(\alpha_s)=-2\alpha_s\sum_{n\ge0}\beta_n\Big(\frac{\alpha_s}{4\pi}\Big)^{n+1}\,,\qquad
\beta_0=\frac{11}{3}C_A-\frac{4}{3}T_Fn_f\,,\qquad
\beta_1=\frac{34}{3}C_A^2-\frac{20}{3}C_AT_Fn_f-4C_FT_Fn_f\,,
\end{equation}
where $n_f$ is the number of light quark flavors. The coefficients of the light-like cusp anomalous dimension are given in (\ref{eq:Gcusp0})-(\ref{eq:Gcusp2}). The coefficients of the anomalous dimensions $\gamma^q$ and $\gamma^Q$, corresponding to the massless quark and heavy quark respectively, are collected as follows~\cite{Moch:2005id,Becher:2006mr,Becher:2009kw,Grozin:2014hna,Grozin:2015kna,Bruser:2019yjk}
\begin{align}
\gamma_0^q={}&-3C_F\,,\label{eq:gq0}\\
\gamma_1^q={}&C_F^2\left(-\frac{3}{2}+2\pi^2-24\zeta_3\right)
+C_FC_A\left(-\frac{961}{54}-\frac{11\pi^2}{6}+26\zeta_3\right)
+C_FT_Fn_f\left(\frac{130}{27}+\frac{2\pi^2}{3}\right)\,,\label{eq:gq1}\\
\gamma_2^q={}&C_F^3\left(-\frac{29}{2}-3\pi^2-\frac{8\pi^4}{5}-68\zeta_3+\frac{16\pi^2}{3}\zeta_3+240\zeta_5\right) \nn\\
&+C_F^2C_A\left(-\frac{151}{4}+\frac{205\pi^2}{9}+\frac{247\pi^4}{135}-\frac{844}{3}\zeta_3-\frac{8\pi^2}{3}\zeta_3-120\zeta_5\right) \nn\\
&+C_FC_A^2\left(-\frac{139345}{2916}-\frac{7163\pi^2}{486}-\frac{83\pi^4}{90}+\frac{3526}{9}\zeta_3-\frac{44\pi^2}{9}\zeta_3-136\zeta_5\right) \nn\\
&+C_F^2T_Fn_f\left(\frac{2953}{27}-\frac{26\pi^2}{9}-\frac{28\pi^4}{27}+\frac{512}{9}\zeta_3\right) \nn\\
&+C_FC_AT_Fn_f\left(-\frac{17318}{729}+\frac{2594\pi^2}{243}+\frac{22\pi^4}{45}-\frac{1928}{27}\zeta_3\right)
+C_FT_F^2n_f^2\left(\frac{9668}{729}-\frac{40\pi^2}{27}-\frac{32}{27}\zeta_3\right)\,,\label{eq:gq2}
\\
\gamma_0^Q={}&-2C_F\,,\label{eq:gQ0}\\
\gamma_1^Q={}&C_FC_A\left(-\frac{98}{9}+\frac{2\pi^2}{3}-4\zeta_3\right)+\frac{40}{9}C_FT_Fn_f\,,\label{eq:gQ1}\\
\gamma_2^Q={}&C_FC_A^2\left(-\frac{343}{9}+\frac{304\pi^2}{27}-\frac{22\pi^4}{45}-\frac{740}{9}\zeta_3-\frac{4\pi^2}{3}\zeta_3+36\zeta_5\right) \nn\\
&+C_F^2T_Fn_f\left(\frac{110}{3}-32\zeta_3\right)
+C_FC_AT_Fn_f\left(\frac{356}{27}-\frac{80\pi^2}{27}+\frac{496}{9}\zeta_3\right)
+\frac{32}{27}C_FT_F^2n_f^2\,.\label{eq:gQ2}
\end{align}

\subsection{C. The result of the radiative jet function at three loops}\label{app:jetfunc}
\begin{figure}
\begin{center}
\includegraphics[width=0.75\textwidth]{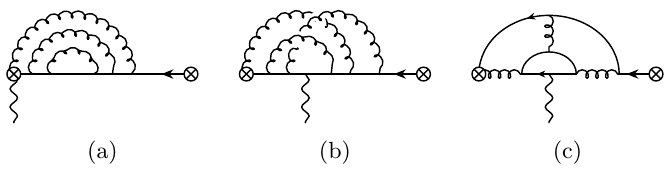}
\end{center}  
\caption{
Sample of Feynman diagrams
contributing to the three-loop radiative jet function:
(a) contribution to $J_A$;
(b) contribution to $J_G$; and 
(c) contribution to $J_G$ from the photon coupled to a light-quark loop .}
\label{fig:jet3l}
\end{figure}

In this section, we present the bare and renormalized radiative jet function in momentum space up to three loops. The bare radiative jet function can be decomposed into two parts, $J^{(0)}(p^2)=J_A^{(0)}(p^2)+J_G^{(0)}(p^2)$, where $J_A^{(0)}$ and $J_G^{(0)}$ are the contributions from the photon and gluon gauge-invariant building blocks in~\eqref{eq:Jdef}, respectively.
We present $J_A^{(0)}$ and $J_G^{(0)}$ contributing to the bare radiative jet function individually, which are given by 
\begin{align}
J_A^{(0)}(p^2)={}&e_q\Bigg\{1+
\frac{Z_\alpha\alpha_s(\mu)}{4\pi}
\left(\frac{-p^2}{\mu^2}\right)^{-\epsilon}
C_F e^{\epsilon\gamma_E}
\frac{\Gamma(1+\epsilon)\Gamma^2(-\epsilon)}{\Gamma(2-2\epsilon)}
\left(4-5\epsilon+\epsilon^2\right)\nn\\
&\qquad+\left(\frac{Z_\alpha\alpha_s(\mu)}{4\pi}\right)^2
\left(\frac{-p^2}{\mu^2}\right)^{-2\epsilon}
C_F\left(C_F K_F^A+C_A K_A^A+T_Fn_f K_{n_f}^A\right)\nn\\
&\qquad+\left(\frac{Z_\alpha\alpha_s(\mu)}{4\pi}\right)^3
\left(\frac{-p^2}{\mu^2}\right)^{-3\epsilon}
\Big(C_F^3 K^A_{FFF}+C_F^2C_A K^A_{FFA}+C_FC_A^2 K^A_{FAA}\nn\\
&\qquad\qquad+C_F^2T_Fn_f K^A_{FFf}+C_FC_AT_Fn_f K^A_{FAf}
+C_F(T_Fn_f)^2 K^A_{Fff}\Big)\Bigg\}
+\mathcal{O}(\alpha_s^4)\,,
\label{eq:JAbare}\\
J_G^{(0)}(p^2)={}&e_q\Bigg\{
\frac{Z_\alpha\alpha_s(\mu)}{4\pi}
\left(\frac{-p^2}{\mu^2}\right)^{-\epsilon}
C_F e^{\epsilon\gamma_E}
\frac{\Gamma(1+\epsilon)\Gamma^2(-\epsilon)}{\Gamma(2-2\epsilon)}
\left(-2+\epsilon-2\epsilon^2\right)\nn\\
&\quad+\left(\frac{Z_\alpha\alpha_s(\mu)}{4\pi}\right)^2
\left(\frac{-p^2}{\mu^2}\right)^{-2\epsilon}
C_F\left(C_F K_F^G+C_A K_A^G+T_Fn_f K_{n_f}^G\right)\nn\\
&\quad+\left(\frac{Z_\alpha\alpha_s(\mu)}{4\pi}\right)^3
\left(\frac{-p^2}{\mu^2}\right)^{-3\epsilon}
\Big(C_F^3 K^G_{FFF}+C_F^2C_A K^G_{FFA}+C_FC_A^2 K^G_{FAA}\nn\\
&\quad\qquad+C_F^2T_Fn_f K^G_{FFf}+C_FC_AT_Fn_f K^G_{FAf}
+C_F(T_Fn_f)^2 K^G_{Fff}\Big)\Bigg\}\nn\\
&+\sum_f e_f\,\left(\frac{Z_\alpha\alpha_s(\mu)}{4\pi}\right)^3
\left(\frac{-p^2}{\mu^2}\right)^{-3\epsilon}
\frac{d^{abc}d^{abc}}{N_c}K^G_{f\gamma}
+\mathcal{O}(\alpha_s^4)\,.
\label{eq:JGbare}
\end{align}
Here $e_{q}$ $(e_{f})$ is the electric charge of the $q$-flavor ($f$-flavor) quark to photon. We omit the ``$-i0$'' prescription in $-p^2-i0$ for brevity.  
We have renormalized the strong coupling in the $\overline{\rm MS}$ scheme, $\alpha_{s,0}=\mu^{2\epsilon}Z_\alpha\alpha_s(\mu)$, with
\begin{equation}
Z_\alpha=1-\frac{\alpha_s}{4\pi}\frac{\beta_0}{\epsilon}
+\Big(\frac{\alpha_s}{4\pi}\Big)^2\Big(\frac{\beta_0^2}{\epsilon^2}-\frac{\beta_1}{2\epsilon}\Big)
+\mathcal{O}(\alpha_s^3)\,.
\end{equation}
The two-loop coefficients $K_F^{A,G}$, $K_A^{A,G}$ and $K_{n_f}^{A,G}$ are given in Appendix~B of~\cite{Liu:2020ydl}. The flavor-singlet contribution to $J_G$ first appears at three loops, as shown in Fig.~\ref{fig:jet3l}\ (c). With ${\rm tr}(T^aT^b)=\delta^{ab}/2$ and $d^{abc}=2{\rm tr}(\{T^a,T^b\}T^c)$, its color factor is
\begin{equation}
\frac{d^{abc}d^{abc}}{N_c}
=\frac{(N_c^2-1)(N_c^2-4)}{N_c^2}\,.
\label{eq:dabccolor}
\end{equation}

For the three-loop coefficients in (\ref{eq:JAbare}) and (\ref{eq:JGbare}) expanded up to $\cO(\epsilon^0)$, we obtain
\begin{align}
K^{A}_{FFF}={}&\frac{32}{3\epsilon^{6}}+\frac{24}{\epsilon^{5}}+\frac{1}{\epsilon^{4}}\Big(74+\frac{8\pi^{2}}{3}\Big)+\frac{1}{\epsilon^{3}}\Big(\frac{407}{2}+2\pi^{2}-112\zeta_3\Big)+\frac{1}{\epsilon^{2}}\Big(\frac{2113}{4}+\frac{11\pi^{2}}{6}-348\zeta_3+\frac{149\pi^{4}}{45}\Big)\nn\\
&+\frac{1}{\epsilon}\Big(\frac{31721}{24}-\frac{29\pi^{2}}{24}-\frac{3976\zeta_3}{3}+\frac{67\pi^{4}}{12}-\frac{316\pi^{2} \zeta_3}{9}-\frac{352\zeta_5}{5}\Big)\nn\\
&+\frac{52025}{16}-\frac{1961\pi^{2}}{144}-\frac{7583\zeta_3}{2}+\frac{6473\pi^{4}}{720}-\frac{101\pi^{2} \zeta_3}{3}-\frac{2492\zeta_5}{5}+\frac{96107\pi^{6}}{34020}+\frac{2200\zeta_3^{2}}{3}+\mathcal{O}(\epsilon).
\label{eq:A-FFF}\\
K^{A}_{FFA}={}&\frac{44}{3\epsilon^{5}}+\frac{1}{\epsilon^{4}}\Big(\frac{565}{9}-\frac{4\pi^{2}}{3}\Big)+\frac{1}{\epsilon^{3}}\Big(\frac{12565}{54}+\frac{2\pi^{2}}{9}-80\zeta_3\Big)+\frac{1}{\epsilon^{2}}\Big(\frac{260833}{324}-\frac{71\pi^{2}}{108}-352\zeta_3-\frac{8\pi^{4}}{5}\Big)\nn\\
&+\frac{1}{\epsilon}\Big(\frac{5153461}{1944}-\frac{9239\pi^{2}}{648}-1473\zeta_3-\frac{43\pi^{4}}{648}+\frac{164\pi^{2} \zeta_3}{9}-488\zeta_5\Big)\nn\\
&+\frac{98365501}{11664}-\frac{294923\pi^{2}}{3888}-\frac{114913\zeta_3}{18}+\frac{87407\pi^{4}}{38880}+\frac{496\pi^{2} \zeta_3}{27}-\frac{40736\zeta_5}{45}-\frac{12685\pi^{6}}{13608}+\frac{3556\zeta_3^{2}}{3}+\mathcal{O}(\epsilon).
\label{eq:A-FFA}\\
K^{A}_{FAA}={}&\frac{484}{81\epsilon^{4}}+\frac{1}{\epsilon^{3}}\Big(\frac{9775}{243}-\frac{88\pi^{2}}{81}\Big)+\frac{1}{\epsilon^{2}}\Big(\frac{48115}{243}-\frac{899\pi^{2}}{243}-\frac{2552\zeta_3}{27}+\frac{44\pi^{4}}{405}\Big)\nn\\
&+\frac{1}{\epsilon}\Big(\frac{3833125}{4374}-\frac{33037\pi^{2}}{2916}-\frac{13420\zeta_3}{27}-\frac{76\pi^{4}}{45}+\frac{88\pi^{2} \zeta_3}{27}+\frac{232\zeta_5}{3}\Big)\nn\\
&+\frac{96985241}{26244}-\frac{367505\pi^{2}}{8748}-\frac{484570\zeta_3}{243}-\frac{8947\pi^{4}}{1080}+\frac{1292\pi^{2} \zeta_3}{27}-\frac{9092\zeta_5}{9}+\frac{5263\pi^{6}}{25515}+\frac{1528\zeta_3^{2}}{9}+\mathcal{O}(\epsilon).
\label{eq:A-FAA}\\
K^{A}_{FFf}={}&-\frac{16}{3\epsilon^{5}}-\frac{188}{9\epsilon^{4}}+\frac{1}{\epsilon^{3}}\Big(-\frac{2110}{27}-\frac{4\pi^{2}}{9}\Big)+\frac{1}{\epsilon^{2}}\Big(-\frac{22355}{81}+\frac{\pi^{2}}{27}+\frac{784\zeta_3}{9}\Big)\nn\\
&+\frac{1}{\epsilon}\Big(-\frac{457367}{486}+\frac{733\pi^{2}}{162}+\frac{10940\zeta_3}{27}-\frac{539\pi^{4}}{810}\Big)\nn\\
&-\frac{9180623}{2916}+\frac{24793\pi^{2}}{972}+\frac{149194\zeta_3}{81}-\frac{15829\pi^{4}}{9720}+\frac{292\pi^{2} \zeta_3}{27}+\frac{1168\zeta_5}{15}+\mathcal{O}(\epsilon).
\label{eq:A-FFf}\\
K^{A}_{FAf}={}&-\frac{352}{81\epsilon^{4}}+\frac{1}{\epsilon^{3}}\Big(-\frac{6640}{243}+\frac{32\pi^{2}}{81}\Big)+\frac{1}{\epsilon^{2}}\Big(-\frac{31514}{243}+\frac{424\pi^{2}}{243}+\frac{736\zeta_3}{27}\Big)\nn\\
&+\frac{1}{\epsilon}\Big(-\frac{1224734}{2187}+\frac{5156\pi^{2}}{729}+\frac{16048\zeta_3}{81}+\frac{68\pi^{4}}{135}\Big)\nn\\
&-\frac{15200537}{6561}+\frac{128693\pi^{2}}{4374}+\frac{75724\zeta_3}{81}+\frac{1307\pi^{4}}{405}-\frac{520\pi^{2} \zeta_3}{27}+\frac{1088\zeta_5}{3}+\mathcal{O}(\epsilon).
\label{eq:A-FAf}\\
K^{A}_{Fff}={}&\frac{64}{81\epsilon^{4}}+\frac{1072}{243\epsilon^{3}}+\frac{1}{\epsilon^{2}}\Big(\frac{544}{27}-\frac{16\pi^{2}}{81}\Big)+\frac{1}{\epsilon}\Big(\frac{186656}{2187}-\frac{268\pi^{2}}{243}-\frac{1472\zeta_3}{81}\Big)\nn\\
&+\frac{2294912}{6561}-\frac{136\pi^{2}}{27}-\frac{24656\zeta_3}{243}-\frac{22\pi^{4}}{81}+\mathcal{O}(\epsilon).
\label{eq:A-Fff}\\
K^{G}_{FFF}={}&-\frac{28}{3\epsilon^{6}}-\frac{24}{\epsilon^{5}}+\frac{1}{\epsilon^{4}}\Big(-76-3\pi^{2}\Big)+\frac{1}{\epsilon^{3}}\Big(-\frac{415}{2}-3\pi^{2}+\frac{256\zeta_3}{3}\Big)+\frac{1}{\epsilon^{2}}\Big(-\frac{2133}{4}-\frac{7\pi^{2}}{3}+266\zeta_3-\frac{113\pi^{4}}{40}\Big)\nn\\
&+\frac{1}{\epsilon}\Big(-\frac{31733}{24}+\frac{61\pi^{2}}{24}+\frac{3094\zeta_3}{3}-\frac{317\pi^{4}}{45}+\frac{112\pi^{2} \zeta_3}{3}+\frac{4984\zeta_5}{15}\Big)\nn\\
&-\frac{157015}{48}+\frac{2681\pi^{2}}{144}+\frac{14389\zeta_3}{6}-\frac{2023\pi^{4}}{120}+\frac{181\pi^{2} \zeta_3}{2}+\frac{4562\zeta_5}{5}-\frac{174827\pi^{6}}{90720}-250\zeta_3^{2}+\mathcal{O}(\epsilon).
\label{eq:G-FFF}\\
K^{G}_{FFA}={}&-\frac{11}{\epsilon^{5}}+\frac{1}{\epsilon^{4}}\Big(-\frac{166}{3}+\pi^{2}\Big)+\frac{1}{\epsilon^{3}}\Big(-\frac{2042}{9}-\frac{41\pi^{2}}{36}+66\zeta_3\Big)+\frac{1}{\epsilon^{2}}\Big(-\frac{88669}{108}-\frac{112\pi^{2}}{27}+311\zeta_3+\frac{23\pi^{4}}{20}\Big)\nn\\
&+\frac{1}{\epsilon}\Big(-\frac{1781851}{648}+\frac{379\pi^{2}}{81}+1217\zeta_3+\frac{33853\pi^{4}}{12960}-\frac{103\pi^{2} \zeta_3}{18}+\frac{638\zeta_5}{3}\Big)\nn\\
&-\frac{33961867}{3888}+\frac{284873\pi^{2}}{3888}+\frac{46258\zeta_3}{9}+\frac{15619\pi^{4}}{19440}-\frac{1541\pi^{2} \zeta_3}{108}+\frac{86467\zeta_5}{45}+\frac{62701\pi^{6}}{272160}-952\zeta_3^{2}+\mathcal{O}(\epsilon).
\label{eq:G-FFA}\\
K^{G}_{FAA}={}&-\frac{242}{81\epsilon^{4}}+\frac{1}{\epsilon^{3}}\Big(-\frac{6521}{243}+\frac{44\pi^{2}}{81}\Big)+\frac{1}{\epsilon^{2}}\Big(-\frac{40289}{243}+\frac{899\pi^{2}}{486}+\frac{1672\zeta_3}{27}-\frac{22\pi^{4}}{405}\Big)\nn\\
&+\frac{1}{\epsilon}\Big(-\frac{1870564}{2187}+\frac{9755\pi^{2}}{2916}+\frac{12590\zeta_3}{27}+\frac{103\pi^{4}}{810}-\frac{44\pi^{2} \zeta_3}{27}-\frac{136\zeta_5}{3}\Big)\nn\\
&-\frac{52268375}{13122}+\frac{58727\pi^{2}}{4374}+\frac{1164763\zeta_3}{486}+\frac{121087\pi^{4}}{19440}-\frac{434\pi^{2} \zeta_3}{27}-\frac{1508\zeta_5}{9}+\frac{613\pi^{6}}{25515}-\frac{1244\zeta_3^{2}}{9}+\mathcal{O}(\epsilon).
\label{eq:G-FAA}\\
K^{G}_{FFf}={}&\frac{4}{\epsilon^{5}}+\frac{56}{3\epsilon^{4}}+\frac{1}{\epsilon^{3}}\Big(\frac{692}{9}+\frac{7\pi^{2}}{9}\Big)+\frac{1}{\epsilon^{2}}\Big(\frac{7561}{27}+\frac{38\pi^{2}}{27}-\frac{500\zeta_3}{9}\Big)+\frac{1}{\epsilon}\Big(\frac{155959}{162}-\frac{167\pi^{2}}{81}-\frac{6712\zeta_3}{27}+\frac{53\pi^{4}}{648}\Big)\nn\\
&+\frac{3134647}{972}-\frac{23485\pi^{2}}{972}-\frac{87584\zeta_3}{81}+\frac{11719\pi^{4}}{4860}-\frac{497\pi^{2} \zeta_3}{27}-\frac{8116\zeta_5}{15}+\mathcal{O}(\epsilon).
\label{eq:G-FFf}\\
K^{G}_{FAf}={}&\frac{176}{81\epsilon^{4}}+\frac{1}{\epsilon^{3}}\Big(\frac{4508}{243}-\frac{16\pi^{2}}{81}\Big)+\frac{1}{\epsilon^{2}}\Big(\frac{27358}{243}-\frac{212\pi^{2}}{243}-\frac{512\zeta_3}{27}\Big)\nn\\
&+\frac{1}{\epsilon}\Big(\frac{1247974}{2187}-\frac{2497\pi^{2}}{729}-\frac{13928\zeta_3}{81}+\frac{2\pi^{4}}{405}\Big)\nn\\
&+\frac{17120104}{6561}-\frac{85447\pi^{2}}{4374}-\frac{82460\zeta_3}{81}-\frac{5851\pi^{4}}{2430}+\frac{268\pi^{2} \zeta_3}{27}+48\zeta_5+\mathcal{O}(\epsilon).
\label{eq:G-FAf}\\
K^{G}_{Fff}={}&-\frac{32}{81\epsilon^{4}}-\frac{752}{243\epsilon^{3}}+\frac{1}{\epsilon^{2}}\Big(-\frac{496}{27}+\frac{8\pi^{2}}{81}\Big)+\frac{1}{\epsilon}\Big(-\frac{199600}{2187}+\frac{188\pi^{2}}{243}+\frac{736\zeta_3}{81}\Big)\nn\\
&-\frac{2710864}{6561}+\frac{124\pi^{2}}{27}+\frac{17296\zeta_3}{243}+\frac{11\pi^{4}}{81}+\mathcal{O}(\epsilon).
\label{eq:G-Fff}\\
K^G_{f\gamma}={}&2+\frac{5\pi^{2}}{6}+\frac{7\zeta_3}{3}-\frac{\pi^{4}}{180}-\frac{40\zeta_5}{3}+\mathcal{O}(\epsilon).
\label{eq:K-fgamma}
\end{align}

The renormalization of the radiative jet function in momentum space is given by the non-local form as follows
\begin{align}
J(p^2,\mu)=\int_0^\infty dx\,Z_J(p^2,xp^2;\mu)\,J^{(0)}(xp^2)\,,  
\label{eq:Jrenorm}
\end{align}
where the renormalization factor $Z_J$ is determined by
\begin{equation}\label{eq:ZJRGE}
\frac{d}{d\ln\mu}Z_J(p^2,xp^2;\mu)=-\int_0^\infty dy\,\gamma_J(p^2,yp^2;\mu)\,Z_J(yp^2,xp^2;\mu)\,.
\end{equation}
The anomalous dimension $\gamma_J$ is derived from the Lange-Neubert kernel based on the RG invariance of the factorization theorem in the leptonic radiative decay. Up to three-loop order, it takes the form
\begin{align}\label{eq:gammaJ}
\gamma_J(p^2,xp^2;\mu)=
\left[\Gamma_{\rm cusp}(\alpha_s)\ln\frac{-p^2}{\mu^2}-\gamma'(\alpha_s)\right]\delta(1-x)  -\Gamma_{\rm cusp}(\alpha_s)\,\Gamma(1,x) 
-\frac{\theta(1-x)}{1-x}\sum_{n=2,3}\left(\frac{\alpha_s}{4\pi}\right)^n h^{(n)}(x)\,,
\end{align}
with $\gamma'=\gamma-\gamma^q-\gamma^Q+\gamma_F$. $\Gamma(1,x)$ is given by~\eqref{eq:LNkernel} with $\omega=1$ and $\omega'=x$.
The coefficients of $\gamma_F(\alpha_s)$ up to three-loop order are~\cite{Broadhurst:1991fz,Chetyrkin:2003vi}
\begin{align}
\gamma_{F,0}={}&-3C_F\,,\label{eq:gF0}\\
\gamma_{F,1}={}&C_F^2\left(\frac{5}{2}-\frac{8\pi^2}{3}\right)
+C_FC_A\left(-\frac{49}{6}+\frac{2\pi^2}{3}\right)
+\frac{10}{3}C_FT_Fn_f\,,\label{eq:gF1}\\
\gamma_{F,2}={}&C_F^3\left(-\frac{37}{2}+\frac{32\pi^2}{3}-\frac{8\pi^4}{9}-36\zeta_3\right)
+C_F^2C_A\left(-\frac{655}{36}-\frac{592\pi^2}{27}-\frac{8\pi^4}{45}+\frac{142}{3}\zeta_3\right)\nn\\
&+C_FC_A^2\left(\frac{1451}{108}+\frac{130\pi^2}{27}-\frac{4\pi^4}{15}-\frac{22}{3}\zeta_3\right)
+C_F^2T_Fn_f\left(\frac{470}{9}+\frac{224\pi^2}{27}-\frac{176}{3}\zeta_3\right)\nn\\
&+C_FC_AT_Fn_f\left(-\frac{512}{27}-\frac{56\pi^2}{27}+\frac{152}{3}\zeta_3\right)
+\frac{140}{27}C_FT_F^2n_f^2\,.\label{eq:gF2}
\end{align}

Finally, we obtain the renormalized radiative jet function in momentum space up to three loops
\begin{align}
J(p^2,\mu)={}&e_q\Bigg\{1+
\frac{\alpha_s}{4\pi}\,C_F\Big(L_p^2-1-\frac{\pi^2}{6}\Big)
+\Big(\frac{\alpha_s}{4\pi}\Big)^2
C_F\Big(C_F k_F+C_A k_A+T_Fn_f k_{n_f}\Big)\nn\\
&+\Big(\frac{\alpha_s}{4\pi}\Big)^3
\Big(C_F^3 k_{FFF}+C_F^2C_A k_{FFA}+C_FC_A^2 k_{FAA}
+C_F^2 T_Fn_f k_{FFf}+C_F C_AT_F n_f k_{FAf}+C_F T_F^2 n_f^2 k_{Fff}\Big)\Bigg\}\nn\\
&+\sum_f e_f\,\Big(\frac{\alpha_s}{4\pi}\Big)^3
\frac{d^{abc}d^{abc}}{N_c}\,K^G_{f\gamma}
+\mathcal{O}(\alpha_s^4)\,,
\label{eq:J-renormalized}
\end{align}
where $L_p\equiv\ln(-p^2/\mu^2)$ and the two-loop coefficients can be found in~\cite{Liu:2020ydl}.
The three-loop coefficients are
\begin{align}
k_{FFF}={}&\frac{1}{6}L_p^{6}+\Big(-\frac{1}{2}-\frac{\pi^{2}}{12}\Big)L_p^{4}+\Big(\pi^{2}+\frac{28\zeta_3}{3}\Big)L_p^{3}+\Big(\frac{3}{2}-\frac{\pi^{2}}{3}-51\zeta_3+\frac{71\pi^{4}}{360}\Big)L_p^{2}\nn\\
&+\Big(-\frac{\pi^{2}}{2}+44\zeta_3-\frac{2\pi^{4}}{15}+\frac{32\pi^{2} \zeta_3}{3}+28\zeta_5\Big)L_p\nn\\
&-\frac{281}{6}-\frac{3\pi^{2}}{2}-\frac{694\zeta_3}{3}-\frac{977\pi^{4}}{360}+\frac{95\pi^{2} \zeta_3}{6}+744\zeta_5-\frac{55771\pi^{6}}{136080}-\frac{290\zeta_3^{2}}{3}\,,\nn\\
k_{FFA}={}&-\frac{11}{9}L_p^{5}+\Big(\frac{67}{9}-\frac{\pi^{2}}{3}\Big)L_p^{4}+\Big(-\frac{272}{27}+\frac{11\pi^{2}}{54}+14\zeta_3\Big)L_p^{3}\nn\\
&+\Big(-\frac{713}{162}-\frac{199\pi^{2}}{36}+\frac{8\zeta_3}{9}-\frac{11\pi^{4}}{60}\Big)L_p^{2}
+\Big(\frac{107}{27}+\frac{1129\pi^{2}}{81}+\frac{2302\zeta_3}{9}-\frac{364\pi^{4}}{135}-\frac{29\pi^{2} \zeta_3}{3}+40\zeta_5\Big)L_p\nn\\
&+\frac{5536}{81}-\frac{2395\pi^{2}}{972}-\frac{5437\zeta_3}{9}+\frac{14015\pi^{4}}{1944}-\frac{194\pi^{2} \zeta_3}{27}-\frac{1502\zeta_5}{9}-\frac{2557\pi^{6}}{68040}+\frac{394\zeta_3^{2}}{3}\,,\nn\\
k_{FAA}={}&\frac{121}{54}L_p^{4}+\Big(-\frac{1780}{81}+\frac{22\pi^{2}}{27}\Big)L_p^{3}+\Big(\frac{13325}{162}-\frac{134\pi^{2}}{27}-44\zeta_3+\frac{11\pi^{4}}{45}\Big)L_p^{2}\nn\\
&+\Big(-\frac{152795}{1458}+\frac{2192\pi^{2}}{243}+\frac{3568\zeta_3}{27}+\frac{319\pi^{4}}{270}-\frac{44\pi^{2} \zeta_3}{9}-96\zeta_5\Big)L_p\nn\\
&-\frac{240323}{26244}-\frac{67655\pi^{2}}{4374}+\frac{57731\zeta_3}{486}-\frac{31579\pi^{4}}{9720}+\frac{275\pi^{2} \zeta_3}{27}-\frac{1954\zeta_5}{9}+\frac{5876\pi^{6}}{25515}+\frac{284\zeta_3^{2}}{9}\,,\nn\\
k_{FFf}={}&\frac{4}{9}L_p^{5}-\frac{20}{9}L_p^{4}+\Big(\frac{100}{27}-\frac{2\pi^{2}}{27}\Big)L_p^{3}+\Big(-\frac{1291}{81}+\frac{17\pi^{2}}{9}+\frac{200\zeta_3}{9}\Big)L_p^{2}\nn\\
&+\Big(\frac{533}{9}-\frac{344\pi^{2}}{81}-\frac{392\zeta_3}{3}+\frac{92\pi^{4}}{135}\Big)L_p
-\frac{37357}{486}+\frac{1537\pi^{2}}{486}+\frac{25052\zeta_3}{81}-\frac{3881\pi^{4}}{2430}+\frac{4\pi^{2} \zeta_3}{27}-\frac{256\zeta_5}{3}\,,\nn\\
k_{FAf}={}&-\frac{44}{27}L_p^{4}+\Big(\frac{1156}{81}-\frac{8\pi^{2}}{27}\Big)L_p^{3}+\Big(-\frac{3310}{81}+\frac{40\pi^{2}}{27}\Big)L_p^{2}
+\Big(\frac{11974}{729}-\frac{628\pi^{2}}{243}-\frac{104\zeta_3}{9}-\frac{34\pi^{4}}{135}\Big)L_p\nn\\
&+\frac{457499}{6561}+\frac{8321\pi^{2}}{2187}-\frac{5900\zeta_3}{81}+\frac{194\pi^{4}}{243}-\frac{40\pi^{2} \zeta_3}{27}+\frac{184\zeta_5}{3}\,,\nn\\
k_{Fff}={}&\frac{8}{27}L_p^{4}-\frac{160}{81}L_p^{3}+\frac{256}{81}L_p^{2}+\Big(\frac{4496}{729}-\frac{64\zeta_3}{27}\Big)L_p
-\frac{132056}{6561}+\frac{8\pi^{2}}{243}+\frac{320\zeta_3}{243}+\frac{4\pi^{4}}{135}\,.
\end{align}

Since the flavor-singlet term is finite in $\epsilon$, its contribution to the renormalized radiative jet function is simply given by the bare coefficient $K^G_{f\gamma}$ in~\eqref{eq:K-fgamma}.

\subsection{D. Phenomenological impact of the three-loop corrections to the radiative jet function}
In order to estimate the phenomenological impact of the three-loop corrections to the jet function,
we recall that, as shown in~\eqref{eq:Bgamma}, the decay amplitude is proportional to the convolution
\begin{equation}\label{eq:jetphi}
(J\otimes\phi)(E_\gamma,\mu) = \int_0^\infty \frac{d\omega}{\omega}\,
J(-2E_\gamma\omega,\mu)\,\phi_B^+(\omega,\mu)\,.
\end{equation}
Following~\cite{Beneke:2011nf}, we define the hadronic matrix elements
\begin{align}
\frac{1}{\lambda_B(\mu)} = \int_0^\infty \frac{d\omega}{\omega}\, \phi_B^+(\omega,\mu)
\qquad\mbox{and}\qquad
\sigma_n(\mu) = \lambda_B(\mu) \int_0^\infty \frac{d\omega}{\omega}\,
\ln^n\frac{\mu_0}{\omega}\, \phi_B^+(\omega,\mu),
\quad\mbox{with}\quad \mu_0=1\,{\rm GeV}.
\label{eq:jetphimom}
\end{align}
For $n_f=4$ and $q=u$, the four active flavors $u,d,s,c$ give
$\sum_f e_f/e_q=1$. The convolution of the radiative jet function and the $B$-meson LCDA takes the form
\begin{equation}
\begin{aligned}
\frac{\lambda_B(\mu_j)}{e_q}(J\otimes\phi)\simeq & \, 1
+\frac{\alpha_s(\mu_j)}{4\pi}\,\left(\frac{4}{3}\sigma_2-\frac{4}{3}-\frac{2\pi^2}{9}\right) 
+ \left(\frac{\alpha_s(\mu_j)}{4\pi}\right)^2
\left(\frac{8}{9}\sigma_4+\frac{100}{27}\sigma_3
+5.990\,\sigma_2 - 55.73\,\sigma_1 - 64.86\right)\\
&\quad
+\left(\frac{\alpha_s(\mu_j)}{4\pi}\right)^3
\left(\frac{32}{81}\sigma_6 + \frac{400}{81}\sigma_5
+26.55\,\sigma_4 - 20.35\,\sigma_3
-546.2\,\sigma_2 - 1639\,\sigma_1-1589\right)
\,,
\end{aligned}
\label{eq:jetlogmom}
\end{equation}
with $\sigma_n\equiv\sigma_n(\mu_j)$. Here the flavor-singlet contribution has been included in the three-loop constant term.

To determine the numeric values of the logarithmic moments $\sigma_n$, we use the reconstructed LCDA from the lattice QCD calculations in~\cite{LPC:2026vyv}. The input parameters are chosen as
\begin{equation}
\mu=\mu_j=m_D=1.87\,\mathrm{GeV},\qquad E_\gamma=\frac{\mu_j^2}{2\mu_0}=1.74845\,\mathrm{GeV}\,,
\label{eq:jetmu}
\end{equation}
so that $L_p=\ln(2E_\gamma\omega/\mu_j^2)=-\ln(\mu_0/\omega)$ and $\alpha_s(\mu_j) \simeq 0.31$\,.
The numeric values of the moments are then given by
\begin{equation}
\begin{aligned}
\sigma_1(\mu_j)&=1.492\,,&\qquad\sigma_2(\mu_j)&=3.842\,,\\
\sigma_3(\mu_j)&=13.15\,,&\qquad\sigma_4(\mu_j)&=55.42\,,\\
\sigma_5(\mu_j)&=288.6\,,&\qquad\sigma_6(\mu_j)&=1755\,.
\end{aligned}
\label{eq:inputmoments}
\end{equation}
Finally, the numerical result for the convolution of the radiative jet function and the $B$-meson LCDA up to three loops is given by
\begin{equation}
\frac{J\otimes\phi}{e_q}\simeq
\frac{1}{\lambda_B(\mu_j)}
\left[ 1 + \left(\frac{\alpha_s(\mu_j)}{0.31}\right)0.03938
- \left(\frac{\alpha_s(\mu_j)}{0.31}\right)^2 0.01646
- \left(\frac{\alpha_s(\mu_j)}{0.31}\right)^3 0.04221
\right]
\,.
\label{eq:jetphen3lnum0}
\end{equation}

For the phenomenological analysis of the leptonic radiative decay $B\to\gamma\ell\nu$ in Fig.~\ref{fig:phenobr}, the partial branching fraction with a photon-energy cut $E_\gamma>E_{\rm cut}$ is given by~\cite{Beneke:2011nf}
\begin{align}
\Delta\mathcal{B}(E_\gamma>E_{\rm cut})=
&\tau_{B^-}\frac{\alpha_{\rm em}G_F^2|V_{ub}|^2m_B^4}{48\pi^2}\int_{E_{\rm cut}}^{m_B/2}dE_\gamma\,
\left(\frac{2E_\gamma}{m_B}\right)^3\,\left(1-\frac{2E_\gamma}{m_B}\right)\Big[F_V^2(E_\gamma)+F_A^2(E_\gamma)\Big]\,,
\label{eq:DeltaB}
\end{align}
where $E_\gamma$ is the photon energy in the $B$-meson rest frame, $m_B$ and $\tau_{B^-}$ are the mass and lifetime of the $B^-$ meson, $V_{ub}$ is the Cabibbo--Kobayashi--Maskawa (CKM) matrix element, and $\alpha_{\rm em}$ and $G_F$ are the fine-structure constant and the Fermi constant, respectively. The vector and axial-vector form factors $F_V$ and $F_A$ parametrize the hadronic matrix element of the weak current with a real photon~\cite{Beneke:2011nf}, which can be factorized at leading power in $\Lambda_{\rm QCD}/m_b$ as follows
\begin{equation}
F_V^{\rm LP}=F_A^{\rm LP}=\frac{m_B f_B}{2E_\gamma}\,
\frac{H(E_\gamma,m_b,\mu_h)}{K(m_b,\mu_h)}\,
U_H(E_\gamma;\mu_h,\mu_j)
\int_0^\infty\frac{d\omega}{\omega}\,J(-2E_\gamma\omega,\mu_j)\,\phi_+^B(\omega,\mu_j)\,.
\label{eq:FVALP}
\end{equation}
Here $H(E_\gamma,m_b,\mu_h)$ is the hard matching coefficient of the heavy-to-light current~\cite{Fael:2024vko}, $K(m_b,\mu_h)$ relates the QCD decay constant $f_B$ to the HQET decay constant through $f_B\sqrt{m_B}=K(m_b,\mu)\,F(\mu)$~\cite{Bekavac:2009zc}, and $U_H(E_\gamma;\mu_h,\mu_j)$ denotes the RG evolution of $H/K$ from the hard scale $\mu_h$ to the jet scale $\mu_j$. The radiative jet function $J$ and the LCDA $\phi_+^B$ are both evaluated at $\mu_j$. The input parameters for the phenomenological analysis are chosen as follows~\cite{Bazavov:2017lyh,Belle-II:2024xwh,ParticleDataGroup:2024cfk} 
\begin{gather}
   G_F = 1.16638\times 10^{-5} {\rm GeV}^{-2}\,,\qquad
   \alpha_{\rm em} = 1/137.036\,,\qquad
   \alpha_s^{(4)}(\mu_j)=0.31\,, \qquad 
   m_b^{\rm pole}=4.8\,\rm{GeV}\,, \nn\\
   f_B=189.4\,\rm{MeV}\,, \qquad
   |V_{ub}|=3.93\times10^{-3}\,, \qquad
   m_B=5.279\,\rm{GeV}\,, \qquad
   \tau_{B^-}=1.638\,\rm{ps}\,.
\end{gather}

\subsection{E. Solution of the three-loop jet evolution equation}

Using the method of~\cite{Liu:2020ydl} to solve the RG equation of the radiative jet function at three loops, the RG evolved radiative jet function takes the following form in momentum space
\begin{equation}
\label{eq:jetevol}
\begin{aligned}
J(p^2,\mu)= & \exp\!\big[ -2S(\mu_j,\mu)-a_{\gamma'}(\mu_j,\mu)
 -2\gamma_Ea_\Gamma(\mu_j,\mu)\big]
\\ &\times\EuScript{J}(\partial_\eta,\mu_j)
 \left(\frac{-p^2}{\mu_j^2}\right)^{\eta+a_\Gamma(\mu_j,\mu)}
 \frac{ \Gamma\big(1-\eta-a_\Gamma(\mu_j,\mu)\big)\Gamma(1+\eta)}
 {\Gamma\big(1+\eta+a_\Gamma(\mu_j,\mu)\big)\Gamma(1-\eta)}\, U_h(\mu_j,\mu,\eta)\Bigg|_{\eta=0}\,,
\end{aligned}
\end{equation}
where the definitions of the Sudakov exponent $S(\mu_j,\mu)$ and the exponent  $a_\gamma(\mu_j,\mu)$ can be found in~\cite{Becher:2007ty}, and 
\begin{align}
  U_h(\mu_j,\mu,\eta) =  \exp\!\Bigg[ -4 C_F\beta_0
 \int_{\alpha_s(\mu_j)}^{\alpha_s(\mu)}
 \frac{d\alpha}{\beta(\alpha)}
 \sum_{n=2}^{3}
 \left(\frac{\alpha}{4\pi}\right)^{n}
 \mathcal H^{(n)}
 \big(\eta+a_\Gamma(\mu_j,\mu_\alpha)\big)
 \Bigg]\,,
\end{align}
with
\begin{equation}
\label{eq:H-three-loop-definition}
 \mathcal H^{(n)}(a)
 =-\frac{1}{4C_F\beta_0}
 \int_0^1 dx\,\frac{x^a h^{(n)}(x)}{1-x},
 \qquad n=2,3\,.
\end{equation}
The derivative of $U_h$ is given by
\begin{equation}
 \frac{d}{d\ln\mu} \ln  U_h(\mu_j,\mu,\eta) 
 =\sum_{n=2}^{3}
 \left(\frac{\alpha_s}{4\pi}\right)^n
 \int_0^1 dx\,\frac{x^{\eta+a_\Gamma(\mu_j,\mu)}h^{(n)}(x)}{1-x},
\end{equation}
which reproduces the $h^{(2)}$ and $h^{(3)}$ contributions
to $dJ/d\ln\mu=-\gamma_J\otimes J$.

The operator $\EuScript{J}(\partial_\eta,\mu_j)$ can be obtained by replacing
$L_p$ of the renormalized jet function in \eqref{eq:J-renormalized} with $\partial_\eta$\,. 
The function $\mathcal H^{(2)}(a)$ is identical to eq.~(53) in~\cite{Liu:2020ydl}.
Inserting \eqref{eq:h3classic} into \eqref{eq:H-three-loop-definition}, we obtain
\begin{equation}
\label{eq:H-three-loop-explicit}
\begin{aligned}
\mathcal H^{(3)}(a)
=-\frac{1}{4\beta_0}&\Bigg\{ 4C_F^2\psi^{(4)}(1+a) +6C_F(\beta_0-3C_F)\psi'''(1+a)
 \\
&+\left[ C_F^2\left(50-\frac{8\pi^2}{3}\right) +8C_F\beta_0+4C_FC_A(3-\pi^2)+2\beta_0^2  \right]\psi''(1+a)
\\
&+\Bigg[
 C_F^2(4\pi^2-6)+4C_FC_A(\pi^2-16)
 -C_F\beta_0\left(10+\frac{4\pi^2}{3}\right)
 -28C_A^2+\frac{4}{3}(19-\pi^2)C_A\beta_0
 +\frac{20}{3}\beta_0^2
 \Bigg]\psi'(1+a)
\\
&+C_F\sum_{n=1}^\infty\Bigg[
 \frac{12C_F}{n}
 \big[\psi'''(a+n)+\psi'''(a+n+1)\big]
\\
&\qquad\quad
 +\frac{8\big[(2C_F+C_A)(\psi(n)+\gamma_E) -3C_F+\beta_0\big]}{n}
 \big[\psi''(a+n)+\psi''(a+n+1)\big] +\frac{16C_F}{n^2}\psi''(a+n+1)
\\
&\qquad\quad
 +\left(
 \frac{8(C_F-C_A)}{n^2}
 \big[\psi(n)+\gamma_E-2n\psi'(n)\big]
 -\frac{24C_F}{n^3}+\frac{4\Gamma_1}{C_F n}
 \right)
 \big[\psi'(a+n)+\psi'(a+n+1)\big]
\\
&\qquad\quad
 +\frac{8(3C_F-\beta_0)}{n^2}\psi'(a+n)
 +\frac{4(C_F-C_A)}{n(a+n)^2}
 \Bigg]\Bigg\}.
\end{aligned}
\end{equation}
Here $\psi^{(m)}$ denotes the $m$th derivative of the digamma function, and $\Gamma_1$ is the coefficient of $\Gamma_{\rm cusp}$ at $\cO(\alpha_s^2)$\,.
In \eqref{eq:jetevol}, the cusp anomalous dimension and the $\beta$ function are needed through four loops, and the remaining anomalous dimensions through three loops.

\end{document}